\documentclass[12pt]{article}
\usepackage{amssymb,amsmath,epsfig}
\allowdisplaybreaks

\begin{document}

\title{\bf Thermodynamics in $f(\mathcal{G},T)$ Gravity}
\author{M. Sharif \thanks {msharif.math@pu.edu.pk}~ and Ayesha Ikram
\thanks{ayeshamaths91@gmail.com}\\
Department of Mathematics, University of the Punjab,\\
Quaid-e-Azam Campus, Lahore-54590, Pakistan.}

\date{}
\maketitle

\begin{abstract}
This paper explores the non-equilibrium behavior of thermodynamics
at the apparent horizon of isotropic and homogeneous universe model
in $f(\mathcal{G},T)$ gravity ($\mathcal{G}$ and $T$ represent the
Gauss-Bonnet invariant and trace of the energy-momentum tensor,
respectively). We construct the corresponding field equations and
analyze the first as well as generalized second law of
thermodynamics in this scenario. It is found that an auxiliary term
corresponding to entropy production appears due to the
non-equilibrium picture of thermodynamics in first law. The
universal condition for the validity of generalized second law of
thermodynamics is also obtained. Finally, we check the validity of
generalized second law of thermodynamics for the reconstructed
$f(\mathcal{G},T)$ models (de Sitter and power-law solutions). We
conclude that this law holds for suitable choices of free
parameters.
\end{abstract}

\section{Introduction}

The discovery of current cosmic accelerated expansion has stimulated
many researchers to explore the cause of this tremendous change in
cosmic history. A mysterious force known as dark energy (DE) is
considered as the basic ingredient responsible for this expanding
phase of the universe. Dark energy has repulsive nature with
negatively large pressure but its complete characteristics are still
unknown. Modified gravity theories are considered as the favorable
and optimistic approaches to unveil the salient features of DE.
These modified theories of gravity are obtained by replacing or
adding curvature invariants as well as their corresponding generic
functions in the Einstein-Hilbert action.

Gauss-Bonnet (GB) invariant is a linear combination of quadratic
invariants of the form
\begin{equation}\nonumber
\mathcal{G}=R^{2}-4R_{\mu\nu}R^{\mu\nu}+R_{\mu\nu\xi\eta}R^{\mu\nu\xi\eta},
\end{equation}
where $R,~R_{\mu\nu}$ and $R_{\mu\nu\xi\eta}$ denote the Ricci
scalar, Ricci and Riemann tensors, respectively. It is the second
order Lovelock scalar with the interesting feature that it is free
from spin-2 ghost instabilities while its dynamical effects do not
appear in four dimensions \cite{1}. The coupling of $\mathcal{G}$
with scalar field or adding generic function $f(\mathcal{G})$ in
geometric part of the Einstein-Hilbert action are the two promising
ways to study the dynamics of $\mathcal{G}$ in four dimensions.
Nojiri and Odintsov \cite{2} presented the second approach as an
alternative for DE referred as $f(\mathcal{G})$ gravity which
explores the fascinating characteristics of late-time cosmological
evolution. This theory is consistent with solar system constraints
and has a quite rich cosmological structure \cite{3}.

The curvature-matter coupling in modified theories has attained much
attention to discuss the cosmic accelerated expansion. Harko et al.
\cite{4} introduced such coupling in $f(R)$ gravity referred as
$f(R,T)$ gravity. Recently, we have established this coupling
between quadratic curvature invariant and matter named as
$f(\mathcal{G},T)$ theory of gravity and found that such coupling
leads to the non-conservation of energy-momentum tensor
$(T_{\mu\nu})$ \cite{5}. Furthermore, the non-geodesic lines of
geometry are followed by massive test particles due to the presence
of extra force while dust particles follow geodesic trajectories.
The stability of Einstein universe is analyzed for both conserved as
well as non-conserved $T_{\mu\nu}$ in this theory \cite{6}. Shamir
and Ahmad \cite{7} applied Noether symmetry approach to construct
some cosmological viable $f(\mathcal{G},T)$ models in the background
of isotropic and homogeneous universe. We have reconstructed the
cosmic evolutionary models corresponding to phantom/non-phantom
epochs, de Sitter universe as well as power-law solution and
analyzed their stability \cite{8}.

The significant connection between gravitation and thermodynamics is
established after the remarkable discovery of black hole (BH)
thermodynamics with Hawking temperature as well as BH entropy
\cite{9}. Jacobson \cite{10} obtained the Einstein field equations
using fundamental relation known as Clausius relation
$dQ=\mathcal{T}d\mathcal{S}$ ($\mathcal{S},~\mathcal{T}$ and $dQ$
represent the entropy, Unruh temperature and energy flux observed by
accelerated observer just inside the horizon, respectively) together
with the proportionality of entropy and horizon area in the context
of Rindler spacetime. Cai and Kim \cite{11} showed that Einstein
field equations can be rewritten in the form of first law of
thermodynamics for isotropic and homogeneous universe with any
spatial curvature parameter. Akbar and Cai \cite{12} found that the
Friedmann equations at the apparent horizon can be written in the
form $dE=\mathcal{T}d\mathcal{S}+Wd\mathcal{V}$ ($E,~\mathcal{V}$
and $W$ are the energy, volume inside the horizon and work density,
respectively) in general relativity (GR), GB gravity and the general
Lovelock theory of gravity. In modified theories, an additional
entropy production term is appeared in Clausius relation that
corresponds to the non-equilibrium behavior of thermodynamics while
no extra term appears in GB gravity, Lovelock gravity and braneworld
gravity \cite{13}.

The generalized second law of thermodynamics (GSLT) has a
significant importance in modified theories of gravity. Wu et al.
\cite{14} derived the universal condition for the validity of GSLT
in modified theories of gravity. Bamba and Geng \cite{15} found that
GSLT in the effective phantom/non-phantom phase is satisfied in
$f(R)$ gravity. Sadjadi \cite{16} studied the second law as well as
GSLT in $f(R,\mathcal{G})$ gravity for de Sitter universe model as
well as power-law solution with the assumption that apparent horizon
is in thermal equilibrium. Bamba and Geng \cite{17} found that GSLT
holds for the FRW universe with the same temperature inside and
outside the apparent horizon in generalized teleparallel theory.
Sharif and Zubair \cite{18} checked the validity of first and second
laws of thermodynamics at the apparent horizon for both equilibrium
as well as non-equilibrium descriptions in $f(R,T)$ gravity and
found that GSLT holds in both phantom as well as non-phantom phases
of the universe. Abdolmaleki and Najafi \cite{19} explored the
validity of GSLT for isotropic and homogeneous universe filled with
radiation and matter surrounded by apparent horizon with Hawking
temperature in $f(\mathcal{G})$ gravity.

In this paper, we investigate the first as well as second law of
thermodynamics at the apparent horizon of FRW model with any spatial
curvature. The paper has the following format. In section
\textbf{2}, we discuss the basic formalism of this gravity while the
laws of thermodynamics are investigated in section \textbf{3}.
Section \textbf{4} is devoted to analyze the validity of GSLT for
reconstructed $f(\mathcal{G},T)$ models corresponding to de Sitter
and power-law solution. The results are summarized in the last
section.

\section{$f(\mathcal{G},T)$ Gravity}

The action of $f(\mathcal{G},T)$ gravity is given by \cite{5}
\begin{equation}\label{1}
\mathcal{I}=\int\sqrt{-g}\left(\frac{R+f(\mathcal{G},T)}{16\pi G}
+\mathcal{L}_{m}\right)d^{4}x,
\end{equation}
where $g,~G$ and $\mathcal{L}_{m}$ represent determinant of the
metric tensor $(g_{\mu\nu})$, gravitational constant and matter
Lagrangian density, respectively. The variation of the action
(\ref{1}) with respect to $g_{\mu\nu}$ gives the fourth-order field
equations as
\begin{eqnarray}\nonumber
R_{\mu\nu}-\frac{1}{2}g_{\mu\nu}R&=&\frac{1}{2}g_{\mu\nu}
f(\mathcal{G},T)-2RR_{\mu\nu}f_{\mathcal{G}}(\mathcal{G},T)
+4R^{\xi}_{\mu}R_{\xi\nu}f_{\mathcal{G}}(\mathcal{G},T)
\\\nonumber&+&4R_{\mu\xi\nu\eta}R^{\xi\eta}f_{\mathcal{G}}
(\mathcal{G},T)-2R_{\mu}^{\xi\eta\delta}R_{\nu\xi\eta\delta}
f_{\mathcal{G}}(\mathcal{G},T)-2Rg_{\mu\nu}\\\nonumber&\times&
\nabla^{2}f_{\mathcal{G}}(\mathcal{G},T)+4R_{\mu\nu}\nabla^{2}
f_{\mathcal{G}}(\mathcal{G},T)+2R\nabla_{\mu}\nabla_{\nu}
f_{\mathcal{G}}(\mathcal{G},T)\\\nonumber&-&4R^{\xi}_{\nu}
\nabla_{\mu}\nabla_{\xi}f_{\mathcal{G}}(\mathcal{G},T)
-4R^{\xi}_{\mu}\nabla_{\nu}\nabla_{\xi}f_{\mathcal{G}}(\mathcal{G},T)
+4g_{\mu\nu}R^{\xi\eta}\\\nonumber&\times&\nabla_{\xi}\nabla_{\eta}
f_{\mathcal{G}}(\mathcal{G},T)-4R_{\mu\xi\nu\eta}\nabla^{\xi}
\nabla^{\eta}f_{\mathcal{G}}(\mathcal{G},T)-(T_{\mu\nu}
+\Theta_{\mu\nu})\\\label{2}&\times&f_{T}(\mathcal{G},T) +8\pi
GT_{\mu\nu},
\end{eqnarray}
where $f_{\mathcal{G}}(\mathcal{G},T)=\frac{\partial
f(\mathcal{G},T)}{\partial\mathcal{G}},
~f_{T}(\mathcal{G},T)=\frac{\partial f(\mathcal{G},T)}{\partial T},
~\nabla^{2}=\nabla_{\mu}\nabla^{\mu}$ ($\nabla_{\mu}$ is a covariant
derivative) and $\Theta_{\mu\nu}$ has the following expression
\cite{20}
\begin{equation}\nonumber
\Theta_{\mu\nu} =-2T_{\mu\nu}+g_{\mu\nu}\mathcal{L}_{m}-2g^{\xi\eta}
\frac{\partial^{2}\mathcal{L}_{m}}{\partial g^{\mu\nu}\partial
g^{\xi\eta}}.
\end{equation}
The variation of $\sqrt{-g}\mathcal{L}_{m}$ with respect to
$g_{\mu\nu}$ yields
\begin{equation}\nonumber
T_{\mu\nu}=g_{\mu\nu}\mathcal{L}_{m}-2\frac{\partial
\mathcal{L}_{m}}{\partial g^{\mu\nu}},
\end{equation}
where we have used that $\mathcal{L}_{m}$ depends only on
$g_{\mu\nu}$.

The covariant derivative of Eq.(\ref{2}) gives
\begin{eqnarray}\nonumber
\nabla^{\mu}T_{\mu\nu}&=&-\frac{f_{T}(\mathcal{G},T)}{8\pi
G-f_{T}(\mathcal{G},T)}\left[\frac{1}{2}g_{\mu\nu}
\nabla^{\mu}T-(\Theta_{\mu\nu}+T_{\mu\nu})\nabla^{\mu} \ln
f_{T}(\mathcal{G},T)\right.\\\label{5}&-&\left.
\nabla^{\mu}\Theta_{\mu\nu}\right].
\end{eqnarray}
The non-zero divergence shows that the conservation law does not
hold in this gravity due to the curvature-matter coupling. The above
equations show that matter Lagrangian density and a generic function
have a significant importance to discuss the dynamics of
curvature-matter coupled theories. The particular forms of
$f(\mathcal{G},T)$ are
\begin{equation}\nonumber
f(\mathcal{G},T)=f_{1}(\mathcal{G})+f_{2}(T),\quad
f(\mathcal{G},T)=f_{1}(\mathcal{G})+f_{2}(\mathcal{G})f_{3}(T),
\end{equation}
where the first choice is considered as correction to
$f(\mathcal{G})$ gravity since it does not involve the direct
non-minimal curvature-matter coupling while the second form implies
direct coupling. Unlike $f(R,T)$ gravity \cite{4}, the choice
$f(\mathcal{G},T)=\mathcal{G}+\lambda T$ ($\lambda$ is an arbitrary
parameter) does not exist in this gravity since $\mathcal{G}$ is a
topological invariant in four dimensions. It is clear from
Eq.(\ref{2}) that the contribution of GB disappears for this
particular choice of the model.

The energy-momentum tensor for perfect fluid as cosmic matter
contents is given by
\begin{equation}\label{6}
T_{\mu\nu}=(\rho+P)v_{\mu}v_{\nu}-Pg_{\mu\nu},
\end{equation}
where $P,~\rho$ and $v_{\mu}$ denote pressure, energy density and
four-velocity, respectively. This four-velocity satisfies the
relations $v^{\xi}v_{\xi}=1$ and $v^{\xi}\nabla_{\nu}v_{\xi}=0$. In
this case, the tensor $\Theta_{\mu\nu}$ with $\mathcal{L}_{m}=-P$
takes the form
\begin{equation}\label{7}
\Theta_{\mu\nu}=-Pg_{\mu\nu}-2T_{\mu\nu}.
\end{equation}
Using Eqs.(\ref{6}) and (\ref{7}), Eq.(\ref{2}) can be written in a
similar form as the Einstein field equations for dust case $(P=0)$
\begin{equation}\label{8}
G_{\mu\nu}=8\pi\widetilde{G}T_{\mu\nu}^{\mathrm{eff}}
=8\pi\widetilde{G}T_{\mu\nu}+T_{\mu\nu}^{\mathrm{(D)}},
\end{equation}
where
\begin{eqnarray}\nonumber
T_{\mu\nu}^{\mathrm{(D)}}&=& \frac{1}{2}g_{\mu\nu}
f(\mathcal{G},T)-2RR_{\mu\nu}f_{\mathcal{G}}(\mathcal{G},T)
+4R^{\xi}_{\mu}R_{\xi\nu}f_{\mathcal{G}}(\mathcal{G},T)
\\\nonumber&+&4R_{\mu\xi\nu\eta}R^{\xi\eta}f_{\mathcal{G}}
(\mathcal{G},T)-2R_{\mu}^{\xi\eta\delta}R_{\nu\xi\eta\delta}
f_{\mathcal{G}}(\mathcal{G},T)-2Rg_{\mu\nu}
\nabla^{2}f_{\mathcal{G}}(\mathcal{G},T)\\\nonumber&+&4R_{\mu\nu}\nabla^{2}
f_{\mathcal{G}}(\mathcal{G},T)+2R\nabla_{\mu}\nabla_{\nu}
f_{\mathcal{G}}(\mathcal{G},T)-4R^{\xi}_{\nu}
\nabla_{\mu}\nabla_{\xi}f_{\mathcal{G}}(\mathcal{G},T)
\\\nonumber&-&4R^{\xi}_{\mu}\nabla_{\nu}\nabla_{\xi}f_{\mathcal{G}}(\mathcal{G},T)
+4g_{\mu\nu}R^{\xi\eta}\nabla_{\xi}\nabla_{\eta}
f_{\mathcal{G}}(\mathcal{G},T)\\\nonumber&-&4R_{\mu\xi\nu\eta}\nabla^{\xi}
\nabla^{\eta}f_{\mathcal{G}}(\mathcal{G},T),\\\nonumber
\widetilde{G}&=&GF,\quad F=1+\frac{f_{T}(\mathcal{G},T)}{8\pi G}.
\end{eqnarray}

The line element for FRW universe model is
\begin{equation}\label{9}
ds^{2}=dt^{2}-\frac{a^{2}(t)}{1-kr^{2}}dr^{2}-\hat{r}^{2}d\theta^{2}
-\hat{r}^{2}\sin^{2}\theta d\phi^{2},
\end{equation}
where $\hat{r}=a(t)r,~a(t)$ and $k$ represent the scale factor
depending on cosmic time and spatial curvature parameter which
corresponds to open $(k=-1)$, closed $(k=1)$ and flat $(k=0)$
geometries of the universe. The GB invariant takes the form
\begin{equation}\nonumber
\mathcal{G}=24(H^{2}+\dot{H})\left(H^{2}+\frac{k}{a^{2}}\right).
\end{equation}
Using Eqs.(\ref{6}), (\ref{8}) and (\ref{9}), we obtain the
following field equations
\begin{eqnarray}\nonumber
3\left(H^{2}+\frac{k}{a^{2}}\right)&=&8\pi
\widetilde{G}\rho+\frac{1}{2}f(\mathcal{G},T)-12(H^{2}+\dot{H})
\left(H^{2}+\frac{k}{a^{2}}\right)\\\nonumber&\times&f_{\mathcal{G}}
(\mathcal{G},T)+12H\left(H^{2}+\frac{k}{a^{2}}\right)
\left(f_{\mathcal{GG}}(\mathcal{G},T)\dot{\mathcal{G}}\right.
\\\label{10}&+&\left.f_{\mathcal{G}T}(\mathcal{G},T)\dot{T}\right),
\\\nonumber-\left(2\dot{H}+3H^{2}+\frac{k}{a^{2}}\right)&=&
-\frac{1}{2}f(\mathcal{G},T)+12(H^{2}+\dot{H})\left(H^{2}+\frac{k}{a^{2}}
\right)f_{\mathcal{G}}(\mathcal{G},T)\\\nonumber&-&8H(H^{2}+\dot{H})
\left(f_{\mathcal{GG}}(\mathcal{G},T)\dot{\mathcal{G}}+f_{\mathcal{G}T}
(\mathcal{G},T)\dot{T}\right)\\\nonumber&-&4\left(H^{2}+\frac{k}{a^{2}}
\right)\left(f_{\mathcal{GGG}}(\mathcal{G},T)\dot{\mathcal{G}}^{2}+2f_{\mathcal{GG}T}
(\mathcal{G},T)\dot{\mathcal{G}}\dot{T}\right.\\\label{11}&+&\left.
f_{\mathcal{G}TT}(\mathcal{G},T)\dot{T}^{2}+f_{\mathcal{GG}}(\mathcal{G},T)
\ddot{\mathcal{G}}+f_{\mathcal{G}T}(\mathcal{G},T)\ddot{T}\right),
\end{eqnarray}
where $H=\dot{a}/a$ is a Hubble parameter and dot represents
derivative with respect to time. We can rewrite the above equations
as
\begin{eqnarray}\label{12}
3\left(H^{2}+\frac{k}{a^{2}}\right)&=&8\pi\widetilde{G}\rho_{\mathrm{_{tot}}}
=8\pi\widetilde{G}(\rho+\rho^{\mathrm{(D)}}),\\\label{13}
-2\left(\dot{H}-\frac{k}{a^{2}}\right)&=&8\pi\widetilde{G}(\rho_{\mathrm{_{tot}}}
+P_{\mathrm{_{tot}}})=8\pi\widetilde{G}(\rho+\rho^{\mathrm{(D)}}+P^{\mathrm{(D)}}),
\end{eqnarray}
where $\rho_{\mathrm{(D)}}$ and $P_{\mathrm{(D)}}$ are dark source
terms given by
\begin{eqnarray}\nonumber
\rho^{\mathrm{(D)}}&=&\frac{1}{8\pi\widetilde{G}F}\left[\frac{1}{2}
f(\mathcal{G},T)-12(H^{2}+\dot{H})\left(H^{2}+\frac{k}{a^{2}}\right)
f_{\mathcal{G}}(\mathcal{G},T)\right.\\\nonumber&+&\left.12H\left(H^{2}
+\frac{k}{a^{2}}\right)\left(f_{\mathcal{GG}}(\mathcal{G},T)\dot{\mathcal{G}}
+f_{\mathcal{G}T}(\mathcal{G},T)\dot{T}\right)\right],\\\nonumber
P^{\mathrm{(D)}}&=&\frac{1}{8\pi\widetilde{G}F}\left[-\frac{1}{2}
f(\mathcal{G},T)+12(H^{2}+\dot{H})\left(H^{2}+\frac{k}{a^{2}}\right)
f_{\mathcal{G}}(\mathcal{G},T)-8H\right.\\\nonumber&\times&\left.(H^{2}
+\dot{H})\left(f_{\mathcal{GG}}(\mathcal{G},T)\dot{\mathcal{G}}
+f_{\mathcal{G}T}(\mathcal{G},T)\dot{T}\right)-4\left(H^{2}+\frac{k}{a^{2}}
\right)\right.\\\nonumber&\times&\left.\left(f_{\mathcal{GGG}}
(\mathcal{G},T)\dot{\mathcal{G}}^{2}+2f_{\mathcal{GG}T}(\mathcal{G},T)\dot{\mathcal{G}}
\dot{T}+f_{\mathcal{G}TT}(\mathcal{G},T)\dot{T}^{2}+f_{\mathcal{GG}}
(\mathcal{G},T)\ddot{\mathcal{G}}\right.\right.\\\nonumber&+&\left.\left.
f_{\mathcal{G}T}(\mathcal{G},T)\ddot{T}\right)\right].
\end{eqnarray}
The continuity equation for Eq.(\ref{9}) becomes
\begin{equation}\label{15}
\dot{\rho}+3H\rho=\frac{-1}{8\pi G+f_{T}(\mathcal{G},T)}
\left[\frac{1}{2}\dot{\rho}f_{T}(\mathcal{G},T)+\rho\left(f_{\mathcal{G}T}
(\mathcal{G},T)\dot{\mathcal{G}}+f_{TT}(\mathcal{G},T)\dot{T}\right)\right].
\end{equation}
The conservation law holds in the absence of curvature-matter
coupling for both $f(\mathcal{G})$ gravity and GR.

\section{Laws of Thermodynamics}

In this section, we study the laws of thermodynamics in the context
of $f(\mathcal{G},T)$ gravity at the apparent horizon of FRW
universe model.

\subsection{First Law}

The first law of thermodynamics is based on the concept that energy
remains conserved in the system but can change from one form to
another. To study this law, we first find the dynamical apparent
horizon evaluated by the relation
\begin{equation}\nonumber
h^{\mu\nu}\partial_{\mu}\hat{r}\partial_{\nu}\hat{r}=0,
\end{equation}
where $h_{\mu\nu}=\mathrm{diag}(1,\frac{-a^{2}}{1-kr^{2}})$ is a
two-dimensional metric. For isotropic and homogeneous universe
model, the above relation gives the radius of apparent horizon as
\begin{equation}\nonumber
\hat{r}_{A}=\left(H^{2}+\frac{k}{a^{2}}\right)^{-\frac{1}{2}}.
\end{equation}
Taking the time derivative of this equation and using Eq.(\ref{13}),
it follows that
\begin{equation}\label{3a}
d\hat{r}_{A}=4\pi
G\left(\rho_{\mathrm{_{tot}}}+P_{\mathrm{_{tot}}}\right)\hat{r}_{A}^{3}HFdt,
\end{equation}
where $d{\hat{r}_{A}}$ represents the infinitesimal change in
apparent horizon radius during the small time interval $dt$.

Bekenstein-Hawking entropy is defined as one fourth of apparent
horizon area $(\mathcal{A}=4\pi\hat{r}_{A}^{2})$ in units of
Newton's gravitational constant \cite{9}. In modified theories of
gravity, the entropy of stationary BH solutions with bifurcate
Killing horizons is a Noether charge entropy also known as Wald
entropy \cite{21}. It depends on the variation of Lagrangian density
$(\mathcal{L})$ with respect to $R_{\mu\nu\xi\eta}$ as \cite{22}
\begin{equation}\label{4a}
\mathcal{S}=-2\pi\oint\frac{\partial\mathcal{L}}{\partial
R_{\mu\nu\xi\eta}}\epsilon_{\xi\eta}\epsilon_{\mu\nu}dV_{n-2}^{2},
\end{equation}
where $dV_{n-2}^{2}$ and $\epsilon_{\xi\eta}$ represent the volume
element on $(n-2)$-dimensional spacelike bifurcation surface
$(\Sigma)$ and binormal vector to $\Sigma$ satisfying the relation
$\epsilon_{\xi\eta}\epsilon^{\xi\eta}=-2$. Brustein et al. \cite{23}
proposed that Wald entropy is equal to quarter of $\mathcal{A}$ in
units of the effective gravitational coupling in modified theories
of gravity. Using these concepts, the Wald entropy in
$f(\mathcal{G},T)$ gravity is given by
\begin{equation}\label{5a}
\mathcal{S}=\frac{\mathcal{A}}{4GF}\left(1-\frac{4}{\hat{r}_{A}^{2}}
f_{\mathcal{G}}(\mathcal{G},T)\right).
\end{equation}
This formula corresponds to $f(\mathcal{G})$ gravity for $F=1$ while
the traditional entropy in GR is obtained for $f_{\mathcal{G}}=0$
\cite{24}. Taking differential of Eq.(\ref{5a}) and using
Eq.(\ref{3a}), we obtain
\begin{eqnarray}\label{6a}
\frac{1}{2\pi\hat{r}_{A}}d\mathcal{S}=4\pi\left(\rho_{\mathrm{_{tot}}}
+P_{\mathrm{_{tot}}}\right)\hat{r}_{A}^{3}Hdt-\frac{2}{\hat{r}_{A}GF}
df_{\mathcal{G}}+\frac{\hat{r}_{A}}{2G}\left(1-\frac{4}{\hat{r}_{A}^{2}}\
f_{\mathcal{G}}\right)d\left(\frac{1}{F}\right).
\end{eqnarray}

The surface gravity $(\kappa_{sg})$ helps to determine temperature
on the apparent horizon as \cite{11}
\begin{equation}\label{7a}
\mathcal{T}=\frac{|\kappa_{sg}|}{2\pi},
\end{equation}
where
\begin{eqnarray}\nonumber
\kappa_{sg}&=&\frac{1}{2\sqrt{-h}}\partial_{\mu}
\left(\sqrt{-h}h^{\mu\nu}\partial_{\nu}\hat{r}\right)
\\\label{8a}&=&\frac{1}{\hat{r}_{A}}\left(1-\frac{\dot{\hat{r}}_{A}}
{2\hat{r}_{A}H}\right)=\frac{1}{2}\hat{r}_{A}\left(\frac{k}{a^{2}}+H^{2}+\dot{H}
\right),
\end{eqnarray}
$h$ is the determinant of $h_{\mu\nu}$. Using
Eqs.(\ref{6a})-(\ref{8a}), we have
\begin{eqnarray}\nonumber
\mathcal{T}d\mathcal{S}&=&4\pi\left(\rho_{\mathrm{_{tot}}}
+P_{\mathrm{_{tot}}}\right)\hat{r}_{A}^{3}Hdt-2\pi\left(\rho_{\mathrm{_{tot}}}
+P_{\mathrm{_{tot}}}\right)\hat{r}_{A}^{2}d\hat{r}_{A}-\frac{4\pi\mathcal{T}}{GF}
df_{\mathcal{G}}\\\label{9a}&+&\frac{\pi}{G}\hat{r}_{A}^{2}\mathcal{T}
\left(1-\frac{4}{\hat{r}_{A}^{2}}f_{\mathcal{G}}\right)d\left(\frac{1}{F}\right).
\end{eqnarray}
The total energy inside the apparent horizon of radius $\hat{r}_{A}$
for FRW universe model is given by
\begin{equation}\nonumber
E=\mathcal{V}\rho_{\mathrm{_{tot}}}=\frac{4}{3}\pi\hat{r}_{A}^{3}
\rho_{\mathrm{_{tot}}}=\frac{3\mathcal{V}}{8\pi\widetilde{G}}\left(H^{2}
+\frac{k}{a^{2}}\right).
\end{equation}
This equation shows that $E$ is directly related to $\hat{r}_{A}$,
so the small displacement $d\hat{r}_{A}$ in horizon radius will
cause the infinitesimal change given by
\begin{eqnarray}\label{11a}
dE=4\pi\rho_{\mathrm{_{tot}}}\hat{r}_{A}^{2}d\hat{r}_{A}-4\pi
\left(\rho_{\mathrm{_{tot}}}
+P_{\mathrm{_{tot}}}\right)\hat{r}_{A}^{3}Hdt+\frac{\hat{r}_{A}}{2G}d
\left(\frac{1}{F}\right).
\end{eqnarray}
Using Eqs.(\ref{9a}) and (\ref{11a}), it follows that
\begin{eqnarray}\nonumber
\mathcal{T}d\mathcal{S}=-dE+Wd\mathcal{V}-\frac{4\pi\mathcal{T}}{GF}d
f_{\mathcal{G}}+\frac{\hat{r}_{A}}{2G}\left[1+2\pi\hat{r}_{A}
\mathcal{T}\left(1-\frac{4}{\hat{r}_{A}^{2}}f_{\mathcal{G}}\right)\right]
d\left(\frac{1}{F}\right),
\end{eqnarray}
where $W=\left(\rho_{\mathrm{_{tot}}} -P_{\mathrm{_{tot}}}\right)/2$
is the work done by the system. The above equation can be written as
\begin{equation}\label{13a}
\mathcal{T}\left(d\mathcal{S}+d_{i}\mathcal{S}\right)=-dE+Wd\mathcal{V},
\end{equation}
where
\begin{equation}\nonumber
d_{i}\mathcal{S}=\frac{4\pi}{GF}df_{\mathcal{G}}-\frac{\hat{r}_{A}}
{2G\mathcal{T}}\left[1+2\pi\hat{r}_{A}\mathcal{T}\left(1-
\frac{4}{\hat{r}_{A}^{2}}f_{\mathcal{G}}\right)\right]
d\left(\frac{1}{F}\right),
\end{equation}
is interpreted as the entropy production term which appears due to
non-equilibrium thermodynamical behavior at the apparent horizon.
The non-equilibrium picture of thermodynamics implies that there is
some energy change inside and outside the apparent horizon. Due to
the presence of this extra term, the field equations do not obey the
universal form of first law of thermodynamics
$dE=\mathcal{T}d\mathcal{S}+Wd\mathcal{V}$ in this gravity. It is
mentioned here that in modified theories, this auxiliary term
usually appears in the first law of thermodynamics while it is
absent in GR, GB gravity and Lovelock gravity \cite{12,13}.

\subsection{Generalized Second Law}

In this section, we discuss the GSLT in $f(\mathcal{G},T)$ gravity
which states that total entropy of the system is not decreasing in
time given by
\begin{equation}\label{1b}
\dot{\mathcal{S}}+\dot{\mathcal{S}}_{\mathrm{_{tot}}}+d_{i}\dot{\mathcal{S}}\geq0,
\end{equation}
where $\mathcal{S}_{\mathrm{_{tot}}}$ is the entropy due to energy
as well as all matter contents inside the horizon and
$d_{i}\dot{\mathcal{S}}=\partial_{t}(d_{i}\mathcal{S})$. The Gibbs
equation relates $\mathcal{S}_{\mathrm{_{tot}}}$ to the total energy
density and pressure as \cite{14}
\begin{equation}\label{2b}
\mathcal{T}_{\mathrm{_{tot}}}d\mathcal{S}_{\mathrm{_{tot}}}
=d(\rho_{\mathrm{_{tot}}}\mathcal{V})+P_{\mathrm{_{tot}}}d\mathcal{V},
\end{equation}
where $\mathcal{T}_{\mathrm{_{tot}}}$ represents total temperature
corresponding to all matter and energy contents inside the horizon
and is not equal to the apparent horizon temperature. We assume
\begin{equation}\nonumber
\mathcal{T}_{\mathrm{_{tot}}}=\zeta\mathcal{T},\quad 0<\zeta<1.
\end{equation}
This proportional relation shows that total temperature inside the
horizon is positive and always smaller than the temperature at the
apparent horizon. Using Eqs.(\ref{13a}) and (\ref{2b}) in
(\ref{1b}), we obtain
\begin{equation}\label{4b}
\dot{\mathcal{S}}+\dot{\mathcal{S}}_{\mathrm{_{tot}}}+d_{i}\dot{\mathcal{S}}
=\left(\frac{24+\hat{r}_{A}^{4}\mathcal{G}}{96\pi
\zeta\hat{r}_{A}}\right)\Upsilon\geq0,
\end{equation}
where
\begin{equation}\nonumber
\Upsilon=(1-\zeta)\mathcal{V}\dot{\rho}_{\mathrm{_{tot}}}+
\left(\rho_{\mathrm{_{tot}}} +P_{\mathrm{_{tot}}}\right)
\left(1-\frac{\zeta}{2}\right)\dot{\mathcal{V}}.
\end{equation}
Using Eqs.(\ref{12}) and (\ref{13}), the GSLT condition takes the
form
\begin{equation}\label{5b}
\left(\frac{24+\hat{r}_{A}^{4}\mathcal{G}}{192\pi\zeta GF}\right)
\hat{r}_{A}^{4}\Xi\geq0,
\end{equation}
where
\begin{equation}\nonumber
\Xi=(2-\zeta)H\left(\dot{H}-\frac{k}{a^{2}}\right)^{2}
+\frac{2(1-\zeta)H}{\hat{r}_{A}}\left(\dot{H}-\frac{k}{a^2}\right)
+\frac{(1-\zeta)F}{\hat{r}_{A}^{4}}\partial_{t}\left(
\frac{1}{F}\right).
\end{equation}
It is seen that GSLT is valid for $\mathcal{G}>0,~F>0$ and $\Xi>0$.
For flat FRW universe model, the conditions
$\mathcal{G}>0,~F>0,~H>0,~\dot{H}>0$ and $\partial_{t}\left(
\frac{1}{F}\right)>0$ must be satisfied to protect the GSLT in
$f(\mathcal{G},T)$ gravity. The equilibrium description of
thermodynamics implies that the temperature inside and at the
horizon are same yielding
\begin{equation}\nonumber
\hat{r}_{A}^{4}H\left(\dot{H}-\frac{k}{a^2}\right)^{2}
\left(\frac{24+\hat{r}_{A}^{4}\mathcal{G}}{192\pi\zeta
GF}\right)\geq0,\quad\zeta=1.
\end{equation}
The validity of GSLT can be obtained for positive values of
$H,~\mathcal{G}$ and $F$.

\section{Validity of GSLT}

Now we check the validity of GSLT for some reconstructed
cosmological models in $f(\mathcal{G},T)$ gravity.

\subsection{de Sitter Universe}

The well-known cosmological de Sitter solution elegantly describes
the evolution of current cosmic expansion. For this model, the
Hubble parameter is constant ($H(t)=H_{0}$) and scale factor grows
exponentially as $a(t)=a_{0}e^{H_{0}t}$. In case of dust fluid,
energy density and GB invariant are given by
\begin{equation}\nonumber
\rho=\rho_{0}e^{-3H_{0}t},\quad\mathcal{G}=24H_{0}^4,
\end{equation}
where $\rho_{0}$ is an integration constant. In this case,
Eq.(\ref{5b}) takes the form
\begin{eqnarray}\nonumber
&&\frac{1+a_{0}^{4}H_{0}^{4}(a_{0}^{2}H_{0}^{2}+ke^{-2H_{0}t})}{\zeta(8\pi
G+f_{T})^{2}}\left[2kH_{0}e^{-2H_{0}t}(b-1)(8\pi
G+f_{T})\right.\\\nonumber&\times&\left.(a_{0}^{2}H_{0}^{4}+ke^{-2H_{0}t})^{-1}
-3\rho_{0}H_{0}e^{-3H_{0}t}(b-1)f_{TT}+k^{2}H_{0}e^{-4H_{0}t}(2-b)
\right.\\\label{2c}&\times&\left.(8\pi
G+f_{T})(a_{0}^{2}H_{0}^{2}+ke^{-2H_{0}t})^{-2}\right]\geq0.
\end{eqnarray}
The reconstructed $f(\mathcal{G},T)$ model for de Sitter universe is
given by \cite{5}
\begin{equation}\nonumber
f(\mathcal{G},T)=c_{1}c_{2}e^{c_{1}\mathcal{G}}T^{-\frac{1}{2}
\left(\frac{1-24c_{1}H_{0}^{4}}{1-36c_{1}H_{0}^{4}}\right)}+
c_{1}c_{2}T^{-\frac{1}{2}}-\frac{16\pi G}{3}T+6H_{0}^{2},
\end{equation}
where $c_{j}$'s $(j=1,2)$ are integration constants and the standard
conservation law is used in the reconstruction technique. The
continuity constraint splits the above model into the following two
$f(\mathcal{G},T)$ forms
\begin{eqnarray}\label{4c}
f_{1}(\mathcal{G},T)&=&\frac{18c_{1}^{2}c_{2}H_{0}^{4}(32c_{1}H_{0}^{4}-1)}
{(1-36c_{1}H_{0}^{2})^{2}}e^{c_{1}\mathcal{G}}T^{-\frac{1}{2}
\left(\frac{1-24c_{1}H_{0}^{4}}{1-36c_{1}H_{0}^{4}}\right)}+6H_{0}^{2}
,\\\nonumber
f_{2}(\mathcal{G},T)&=&\frac{18c_{1}H_{0}^{4}(1-32c_{1}H_{0}^{4})}
{(1-24c_{1}H_{0}^{4})(1-30c_{1}H_{0}^{4})}\left(c_{1}c_{2}T^{-\frac{1}{2}}-
\frac{16\pi G}{3}T\right)+6H_{0}^{2}.\\\label{5c}
\end{eqnarray}
\begin{figure}
\epsfig{file=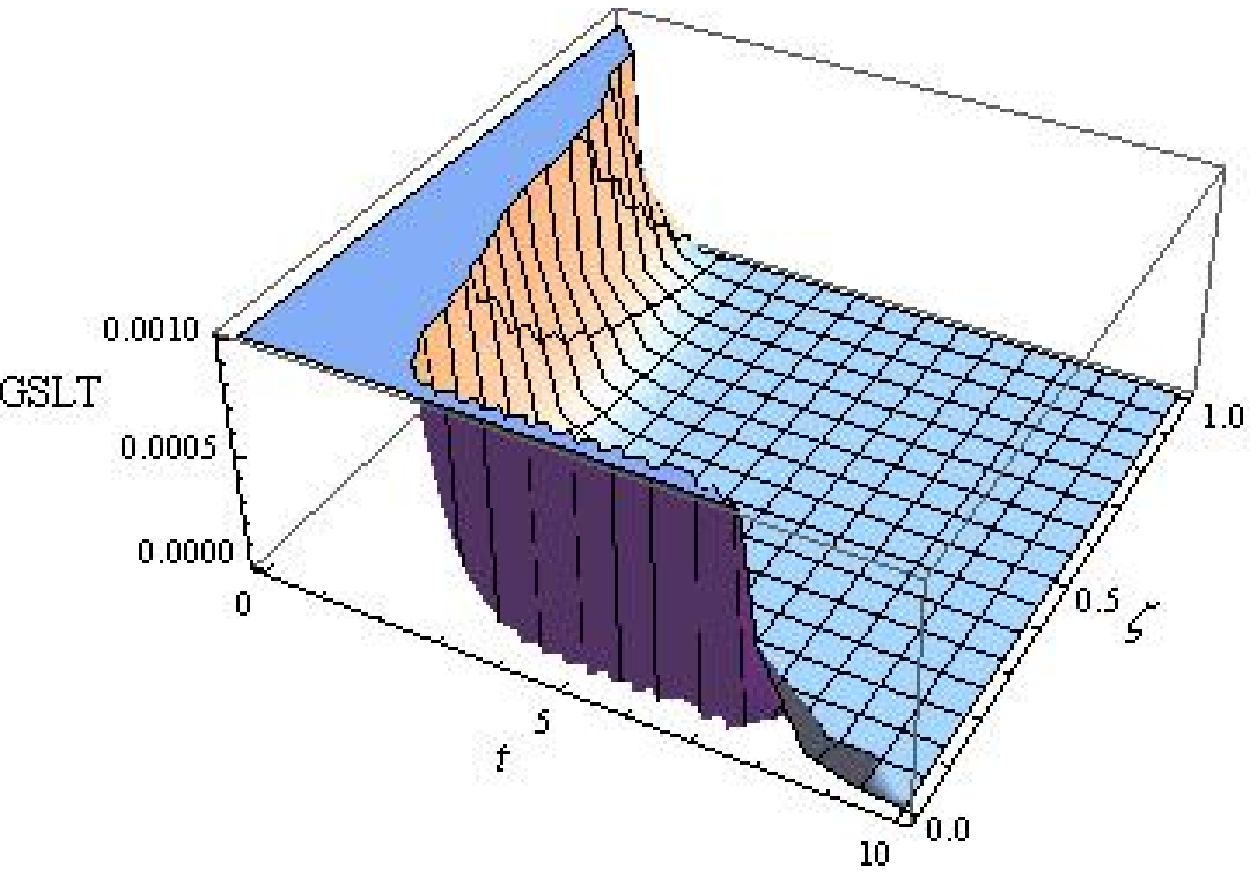, width=0.5\linewidth}\epsfig{file=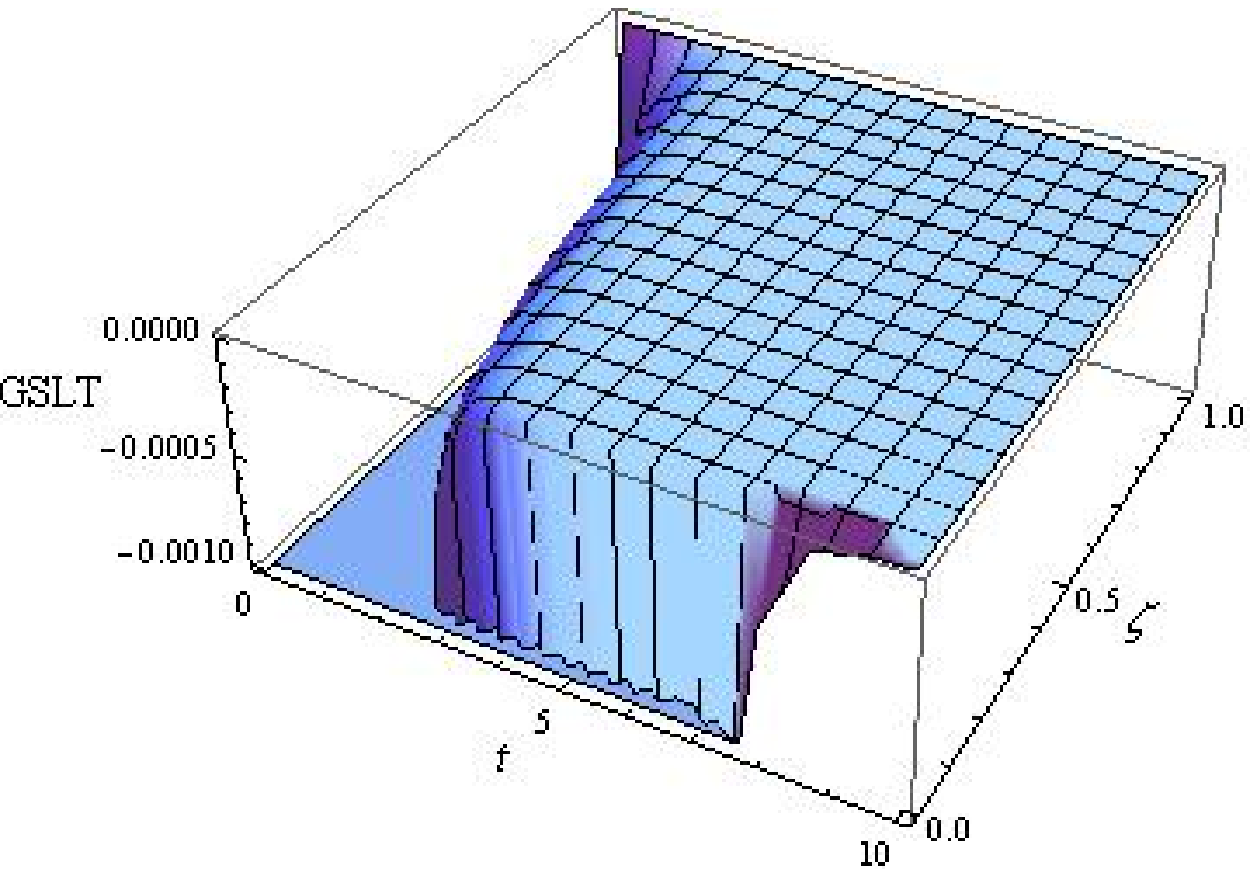,
width=0.5\linewidth}\caption{Validity of GSLT for the model
(\ref{4c}). The left plot is for $c_{1}=c_{2}=1$ and right for
$c_{1}=1$ with $c_{2}=-1$.}
\end{figure}
\begin{figure}
\epsfig{file=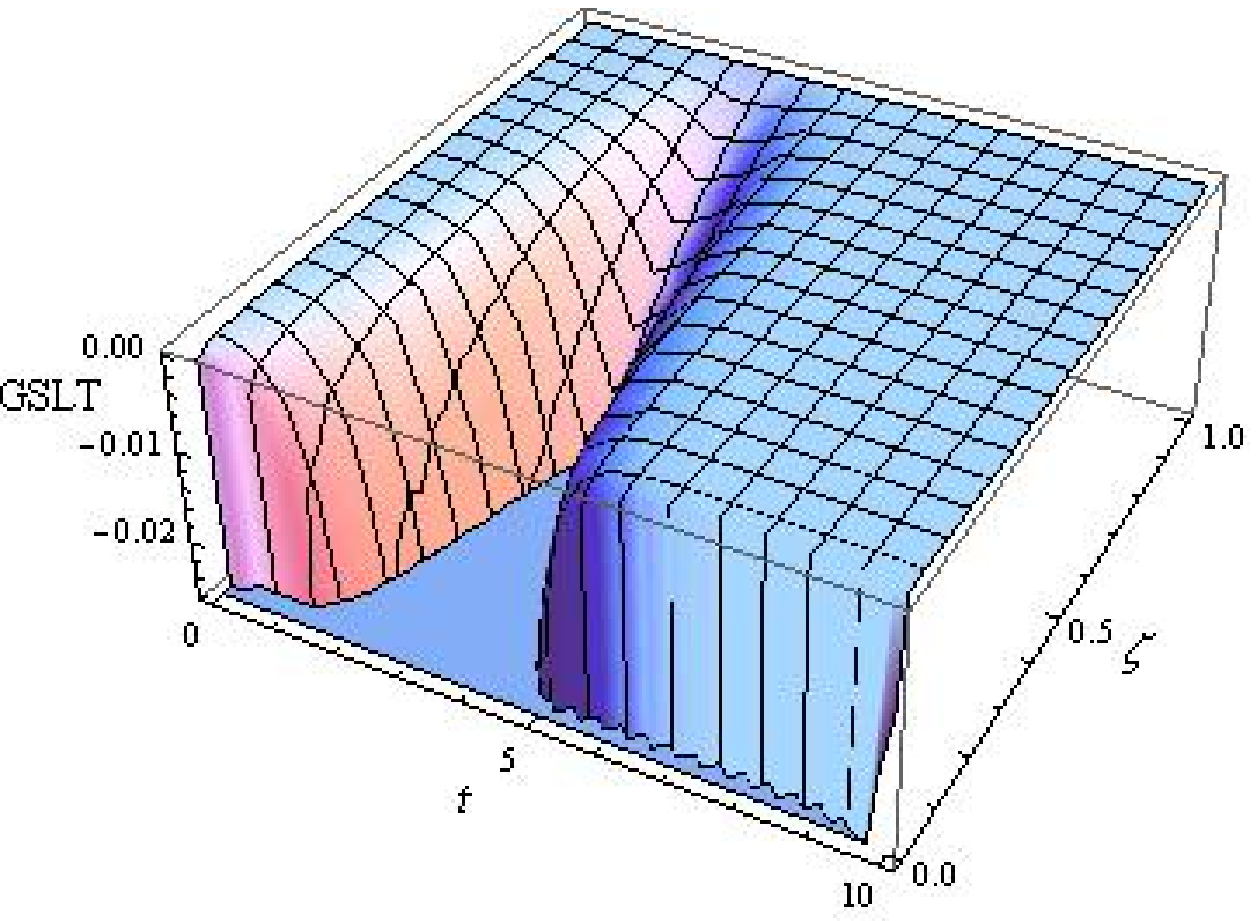, width=0.5\linewidth}\epsfig{file=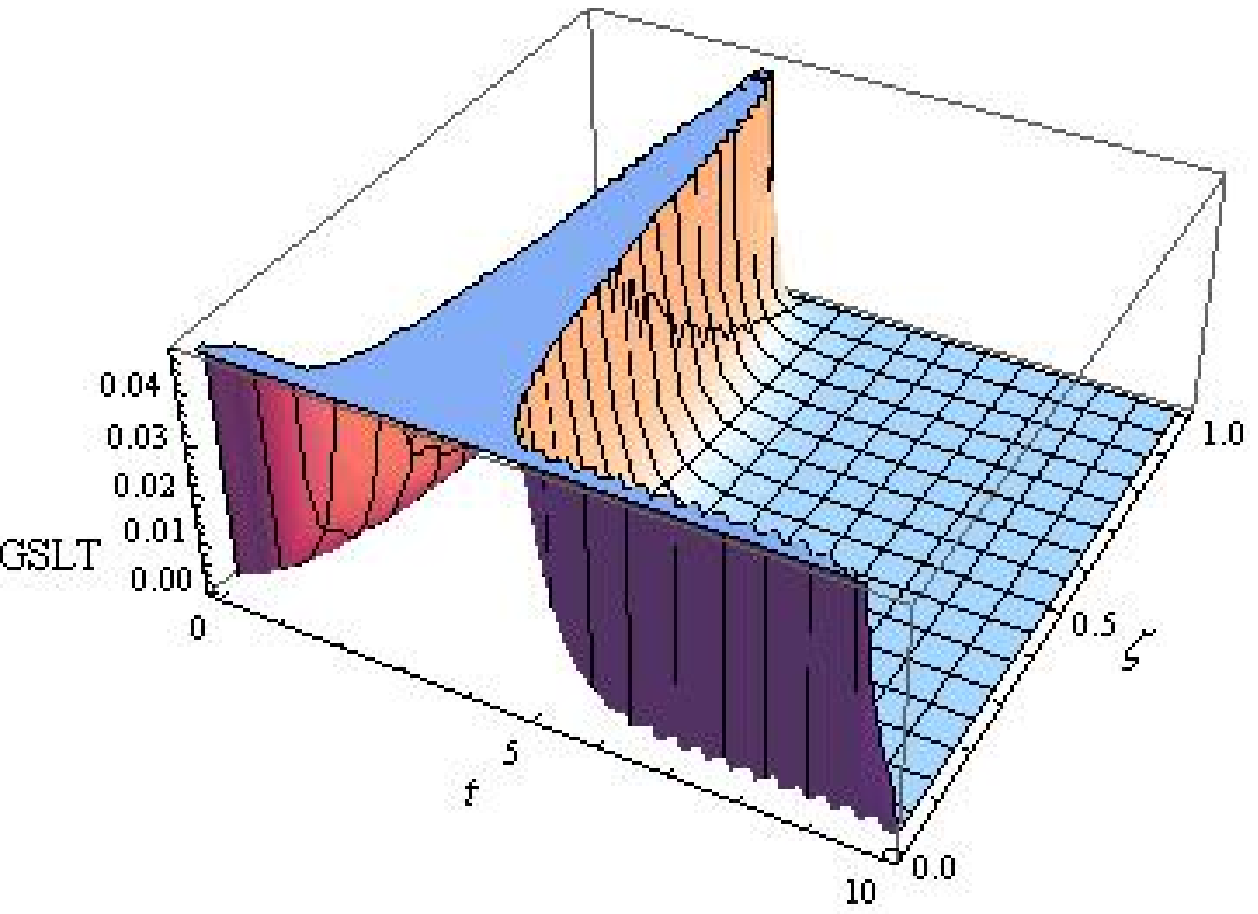,
width=0.5\linewidth}\caption{Validity of GSLT for the model
(\ref{4c}). The left plot is for $c_{1}=-1$ with $c_{2}=1$ and right
for $c_{1}=c_{2}=-1$.}
\end{figure}

Figures \textbf{1} and \textbf{2} show the validity of GSLT for the
model (\ref{4c}) in the background of flat FRW universe model. The
present day value of Hubble parameter is
$H_{0}=(67.8\pm0.9)~\mathrm{kms}^{-1}\mathrm{Mpc}^{-1}$ at the
$68\%$C.L. (C.L. stands for confidence level) which can be
considered as $0.67$ in units of
$100~\mathrm{kms}^{-1}\mathrm{Mpc}^{-1}$ \cite{r1,r2}. The value of
matter density parameter is constrained as $0.308\pm0.012$ with
$68\%$C.L. whereas scale factor at $t_{0}=13.7~\mathrm{Gyr}$ is
$a_{0}=1$ \cite{r1}. For this model, we have four parameters
$c_{1},~c_{2},~\zeta$ and $t$ with fixed values of
$H_{0}=0.67,~a_{0}=1$ and $\rho_{0}=0.3$. Here, we examine the
validity of GSLT against two parameters $\zeta$ and $t$ with four
possible choices of integration constants. For the case
$(c_{1},c_{2})>0$, we find that the validity of GSLT holds for the
considered intervals of $\zeta$ and $t$. Figure \textbf{1} (left)
indicates the validity for $c_{1}=c_{2}=1$ while the right plot
corresponds to the case $c_{1}>0$ and $c_{2}<0$. The left plot of
Figure \textbf{2} shows that the validity of GSLT is not true for
$c_{1}<0$ with $c_{2}>0$ while it satisfies for both negative values
of $(c_{1},c_{2})$ as shown in the right panel. It is found that the
generalized second law holds at all times only for the same
signatures of integration constants.
\begin{figure}
\epsfig{file=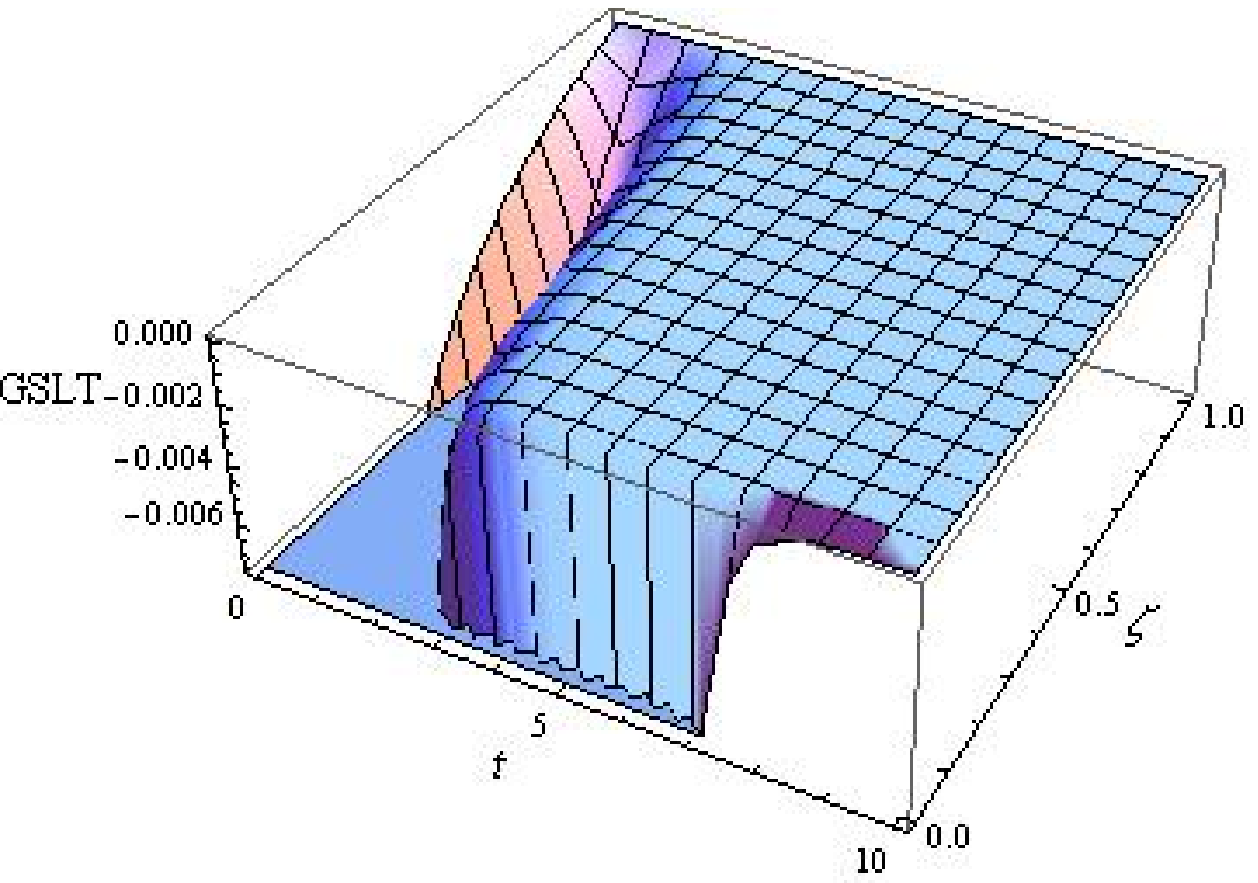, width=0.5\linewidth}\epsfig{file=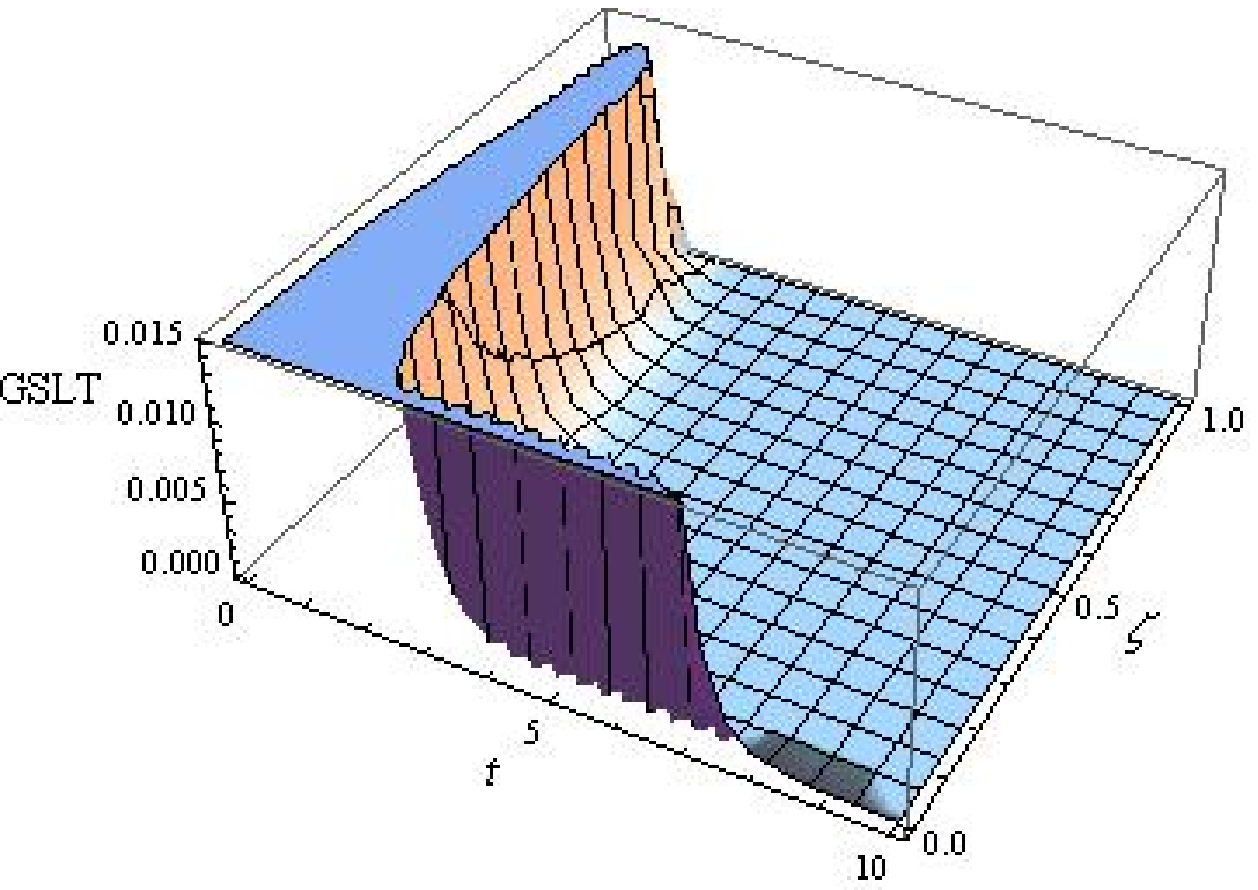,
width=0.5\linewidth}\caption{Validity of GSLT for the model
(\ref{5c}). The left plot is for $c_{1}=c_{2}=1$ and right for
$c_{1}=1$ with $c_{2}=-1$.}
\end{figure}
\begin{figure}
\epsfig{file=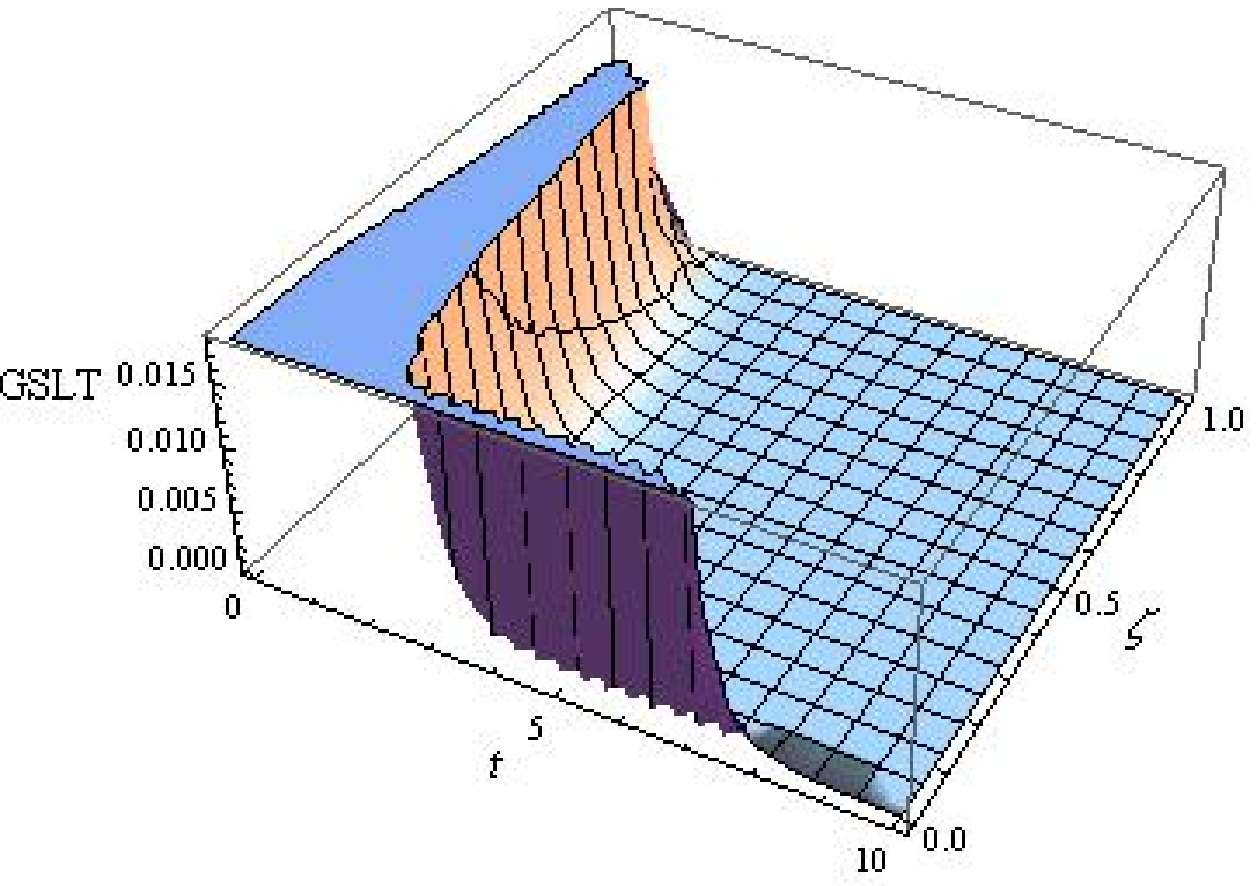, width=0.5\linewidth}\epsfig{file=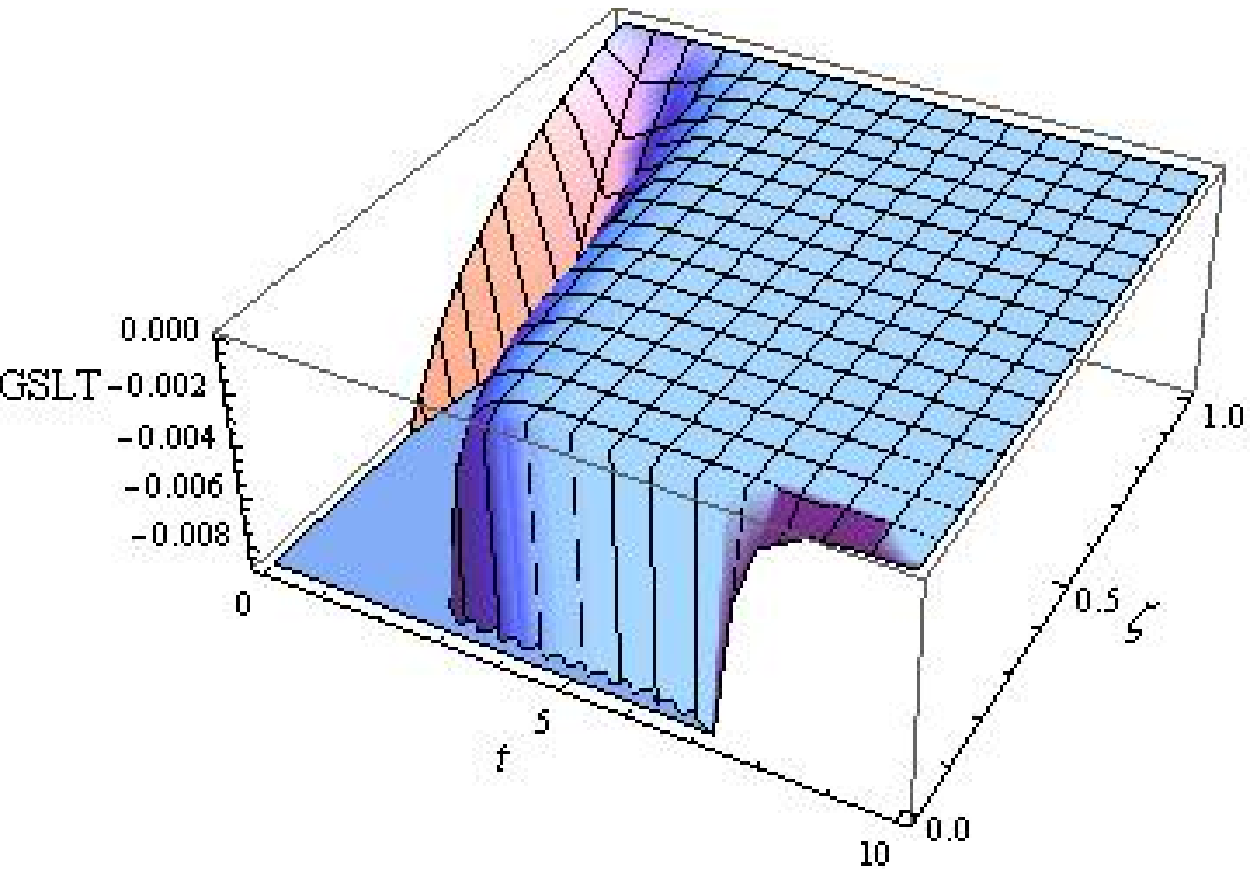,
width=0.5\linewidth}\caption{Validity of GSLT for the model
(\ref{5c}). The left plot is for $c_{1}=-1$ with $c_{2}=1$ and right
for $c_{1}=c_{2}=-1$.}
\end{figure}

The graphical behavior of GSLT for de Sitter $f(\mathcal{G},T)$
model (\ref{5c}) is shown in Figures \textbf{3} and \textbf{4}
against parameters $\zeta$ and $t$. In this case, we have considered
$H_{0}=0.67,~a_{0}=1$ and $\rho_{0}=0.3$ with four possible
signature choices of integration constants $c_{1}$ and $c_{2}$ as in
the previous model. Figure \textbf{3} (right) and Figure \textbf{4}
(left) show that GSLT is true for all considered values of $\zeta$
and $t$ with opposite signatures of $(c_{1},c_{2})$. For model
(\ref{5c}), the choice of same signatures of integration constants
is ruled out since it does not provide feasible region for which
GSLT holds.

\subsection{Power-law Solution}

Power-law solution has remarkable importance in modified theories of
gravity to discuss the decelerated as well as accelerated cosmic
evolutionary phases which are characterized by the scale factor as
\cite{26}
\begin{equation}\label{1p}
a(t)=a_{0}t^{\beta},\quad H=\frac{\beta}{t},\quad\beta>0.
\end{equation}
The accelerated phase of the universe is observed for $\beta>1$
while $0<\beta<1$ covers the decelerated phase including dust
$(\beta=\frac{2}{3})$ as well as radiation $(\beta=\frac{1}{2})$
dominated cosmic epochs. For this scale factor, the energy density
and GB invariant becomes
\begin{equation}\label{2p}
\rho=\rho_{0}t^{-3\beta},\quad\mathcal{G}=24\frac{\beta^3}{t^4}(\beta-1).
\end{equation}
Using Eqs.(\ref{1p}) and (\ref{2p}) in (\ref{5b}), the validity
condition for GSLT takes the form
\begin{eqnarray}\nonumber
&&\frac{1+a_{0}^{4}\beta^{3}(\beta-1)t^{-4}(a_{0}^{2}\beta^{2}t^{-2}+kt^{-2\beta})^{-2}}
{\zeta(8\pi G+f_{T})}\left[\frac{\beta}{t}(2-\zeta)\left(\frac{\beta
a_{0}^{2}t^{-2}+kt^{-2\beta}}
{\beta^{2}a_{0}^{2}t^{-2}+kt^{-2\beta}}\right)^{2}\right.\\\nonumber&-&\left.
2\frac{\beta}{t}(1-\zeta)\left(\frac{\beta
a_{0}^{2}t^{-2}+kt^{-2\beta}}
{\beta^{2}a_{0}^{2}t^{-2}+kt^{-2\beta}}\right)-\frac{\zeta-1}{8\pi
G+f_{T}}\left(96\frac{\beta^{3}}{t^{5}}(\beta-1)f_{\mathcal{G}T}
\right.\right.\\\nonumber&+&\left.\left.3\beta\rho_{0}t^{-4}f_{TT}\right)\right]\geq0.
\end{eqnarray}
The reconstructed $f(\mathcal{G},T)$ model for dust fluid is given
by \cite{5}
\begin{eqnarray}\nonumber
f(\mathcal{G},T)=d_{1}d_{3}T^{d_{2}}
\mathcal{G}^{\frac{1}{4}(\alpha_{1}+\alpha_{2})}+
d_{2}d_{3}T^{d_{2}}\mathcal{G}^{\frac{1}{4}
(\alpha_{1}-\alpha_{2})}+d_{1}d_{2}
T^{\alpha_{3}}+\alpha_{4}T+\alpha_{5}T^{\alpha_{6}},
\end{eqnarray}
where $d_{l}$'s $(l=1,2,3)$ are integration constants and
\begin{eqnarray}\nonumber
\alpha_{1}&=&\frac{1}{2}\left[5-\beta(1+3d_{2})\right],\\\nonumber
\alpha_{2}&=&\left[\frac{3}{4}\beta d_{2}\{3d_{2}\beta+2(\beta-1)
-8\}+\frac{1}{4}(\beta-1)(\beta+7)+4+8d_{2}(\beta-1)\right]^{\frac{1}{2}},
\\\nonumber\alpha_{3}&=&-\frac{1}{2},\quad\alpha_{4}=-\frac{16\pi
G}{3},\quad\alpha_{5}=\left(\frac{18\beta^{3}}{3\beta+4}\right)
\rho_{0}^{-\frac{2}{3\beta}},\quad\alpha_{6}=\frac{2}{3\beta},
\end{eqnarray}
imply that the conservation law holds. The continuity constraint
splits this model into two functions with some additional constants
as
\begin{eqnarray}\label{4p}
f_{1}(\mathcal{G},T)&=&d_{1}d_{3}\gamma_{1}T^{d_{2}}
\mathcal{G}^{\frac{1}{4}(\alpha_{1}+\alpha_{2})}+d_{1}d_{2}
\gamma_{2}T^{\alpha_{3}}+\gamma_{3}T+\gamma_{4}T^{\alpha_{6}},\\\label{5p}
f_{2}(\mathcal{G},T)&=&d_{2}d_{3}\gamma_{5}T^{d_{2}}
\mathcal{G}^{\frac{1}{4}(\alpha_{1}-\alpha_{2})}+d_{1}d_{2}
\gamma_{6}T^{\alpha_{3}}+\gamma_{7}T+\gamma_{8}T^{\alpha_{6}},
\end{eqnarray}
where
\begin{eqnarray}\nonumber
\gamma_{1}&=&1-\frac{\alpha_{7}}{\alpha_{8}},\quad
\gamma_{2}=1-\frac{\alpha_{3}^{2}}{\alpha_{8}},\quad
\gamma_{3}=\alpha_{4}\left(1-\frac{1}{\alpha_{8}}\right),\quad
\gamma_{4}=\alpha_{5}\left(1-\frac{\alpha_{6}^{2}}{\alpha_{8}}\right),
\\\nonumber\gamma_{5}&=&1-\frac{\alpha_{8}}{\alpha_{7}},\quad
\gamma_{6}=1-\frac{\alpha_{3}^{2}}{\alpha_{7}},\quad
\gamma_{7}=\gamma_{4}\left(1-\frac{1}{\alpha_{7}}\right),\quad
\gamma_{8}=\gamma_{5}\left(1-\frac{\alpha_{6}^{2}}{\alpha_{7}}\right),
\\\nonumber\alpha_{7}&=&\frac{d_{2}}{6\beta}\left[6d_{2}\beta
-3\beta+2(\alpha_{1}+\alpha_{2})\right],\quad\alpha_{8}
=\frac{d_{2}}{6\beta}\left[6d_{2}\beta-3\beta+2(\alpha_{1}-\alpha_{2})\right].
\end{eqnarray}
\begin{figure}
\epsfig{file=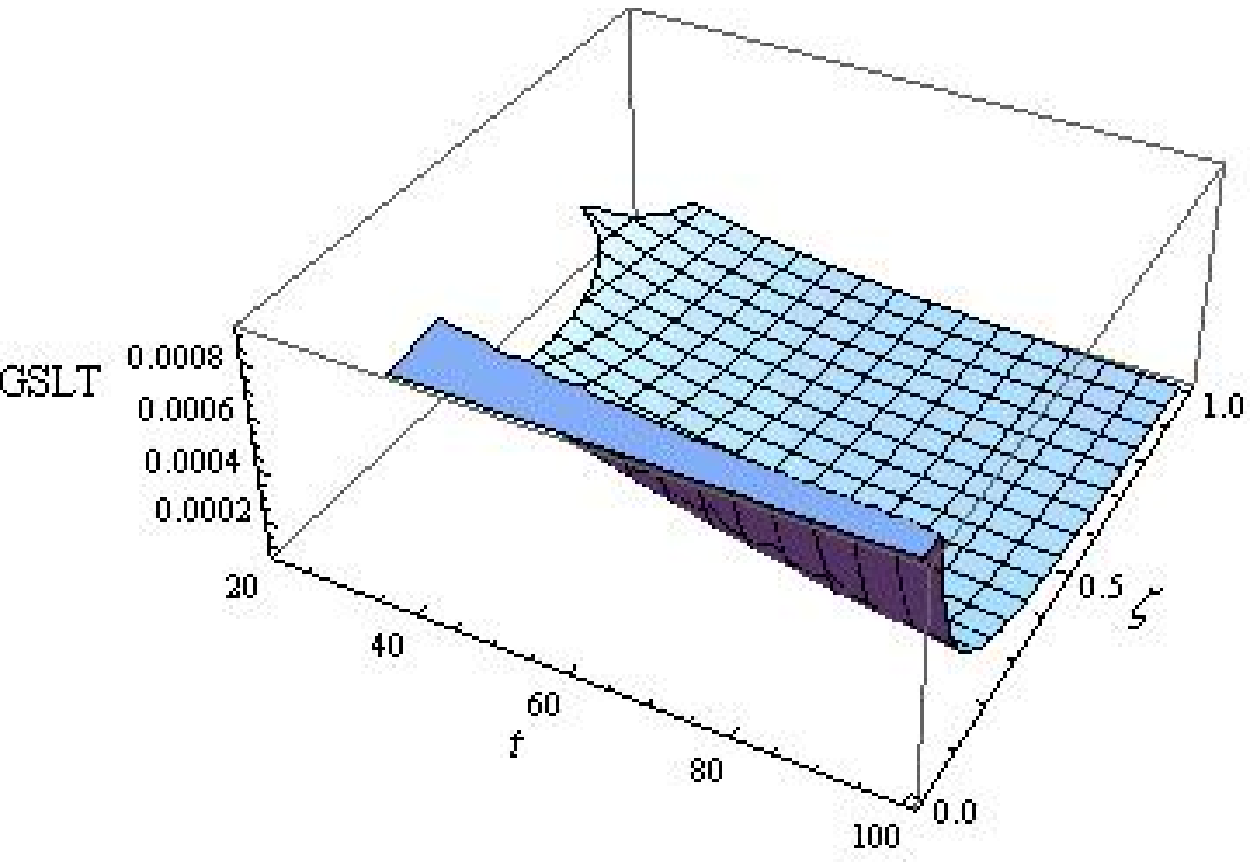, width=0.5\linewidth}\epsfig{file=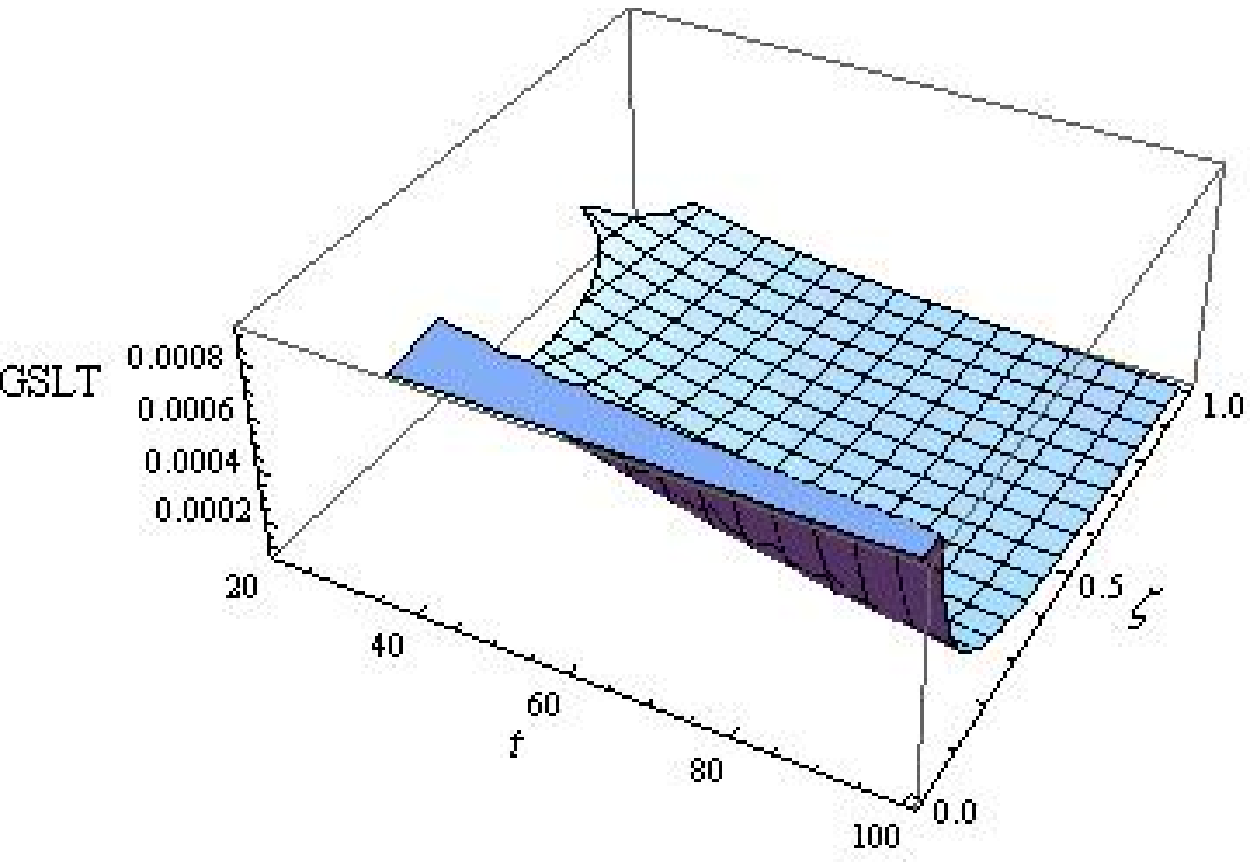,
width=0.5\linewidth}\caption{Validity of GSLT for the model
(\ref{4p}). The left plot is for $d_{1}=d_{3}=1$ and right for
$d_{1}=1$ and $d_{3}=-1$.}
\end{figure}
\begin{figure}
\epsfig{file=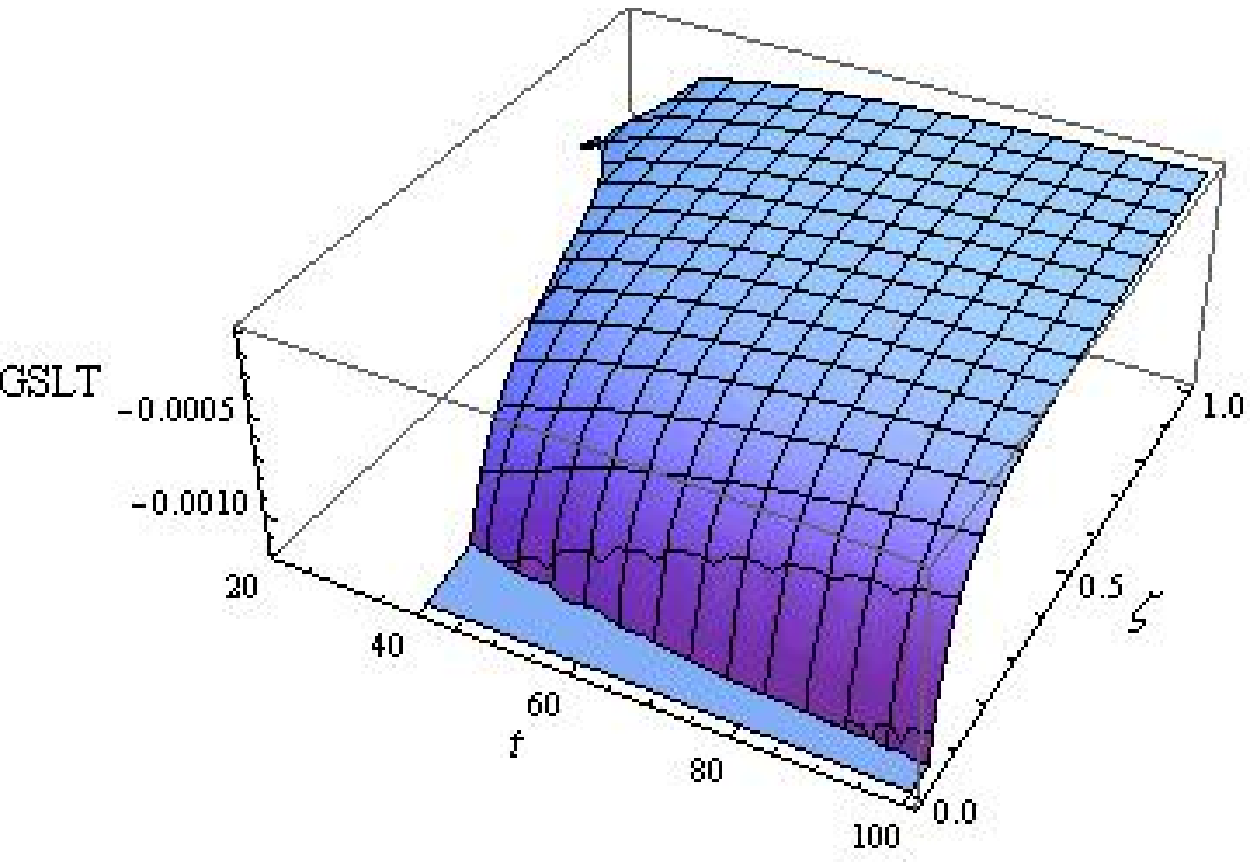, width=0.5\linewidth}\epsfig{file=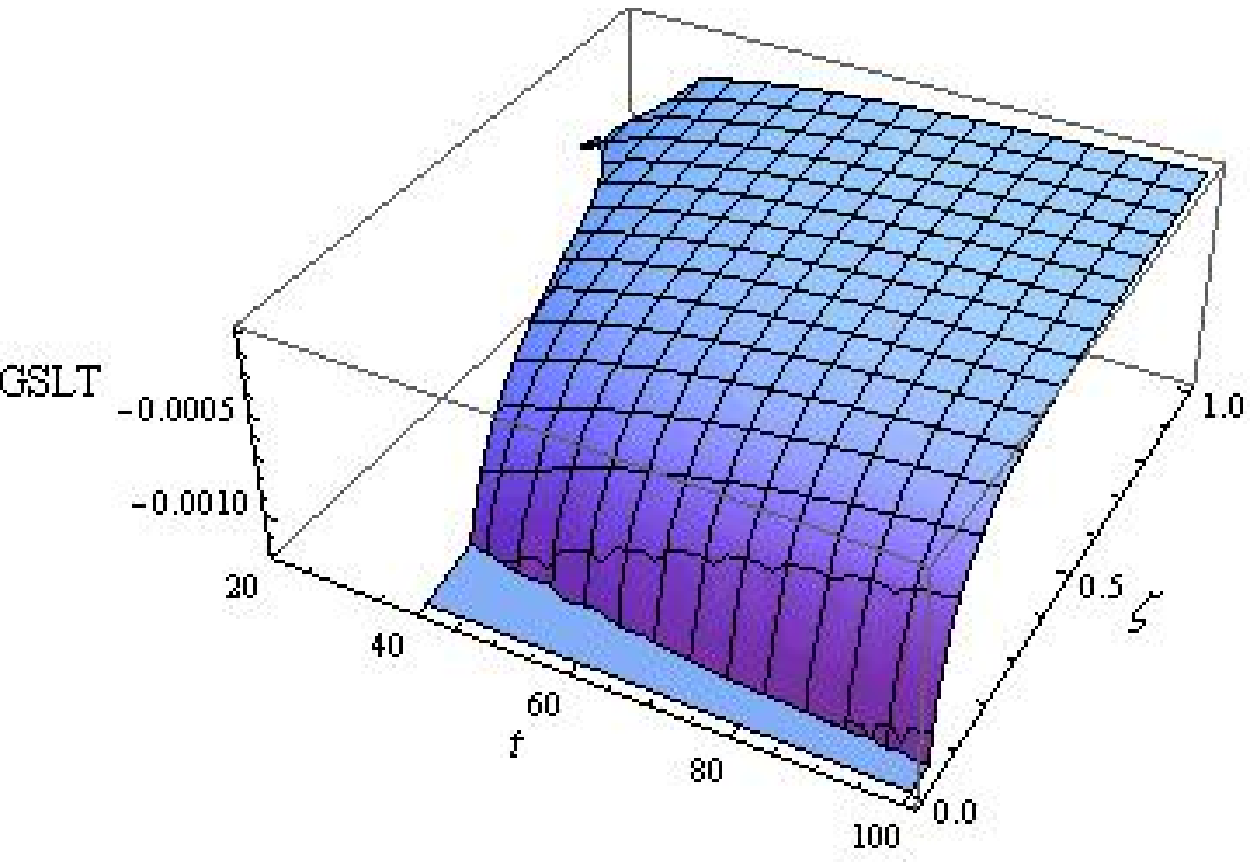,
width=0.5\linewidth}\caption{Validity of GSLT for the model
(\ref{4p}). The left plot is for $d_{1}=-1$ with $d_{3}=1$ and right
for $d_{1}=d_{3}=-1$.}
\end{figure}

The validity of GSLT for models (\ref{4p}) and (\ref{5p}) depend on
five parameters $d_{1},~d_{2},~d_{3},~\zeta$ and $t$. Figure
\textbf{5} and \textbf{6} show the behavior of GSLT for power-law
reconstructed $f(\mathcal{G},T)$ model (\ref{4p}). We examine the
validity against $\zeta$ and $t$ with
$a_{0}=1,~\rho_{0}=0.3,~n=\frac{2}{3}$ and $d_{2}=-1.64285$ while
the behavior of remaining two integration constants $d_{2}$ and
$d_{3}$ are investigated for four possible choices of their
signatures. Figure \textbf{5} shows that GSLT is satisfied for both
cases $d_{3}>0$ and $d_{3}<0$ with $d_{1}>0$ in the considered
interval of parameters $(\zeta,t)$. We also check that the validity
region decreases as the value of integration constants increase
positively as well as negatively. Figure \textbf{6} shows that GSLT
does not hold for model (\ref{4p}) when $d_{3}>0$ and $d_{3}<0$ with
$d_{1}<0$. From both figures, it is found that the signature of
$d_{1}$ has dominant effect on the validity of GSLT as compared to
$d_{3}$.
\begin{figure}
\epsfig{file=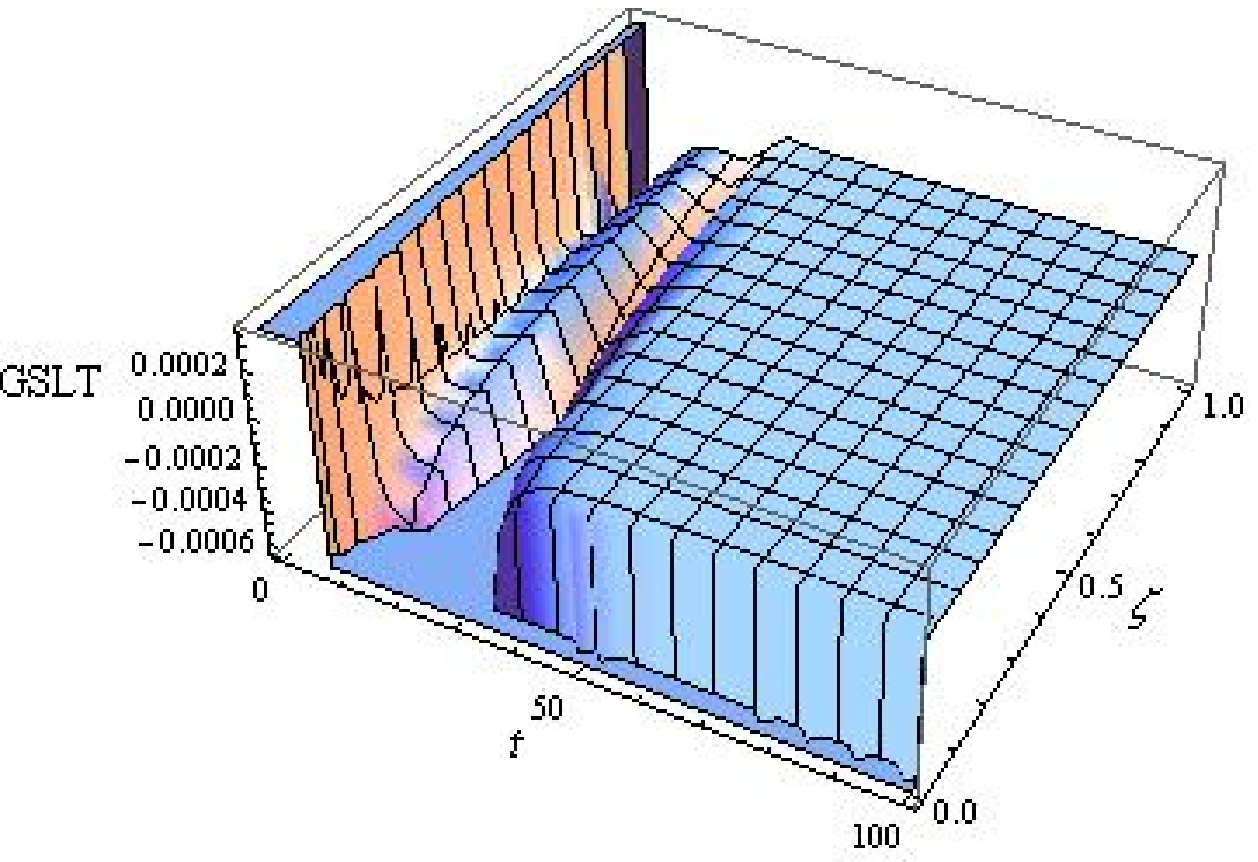, width=0.5\linewidth}\epsfig{file=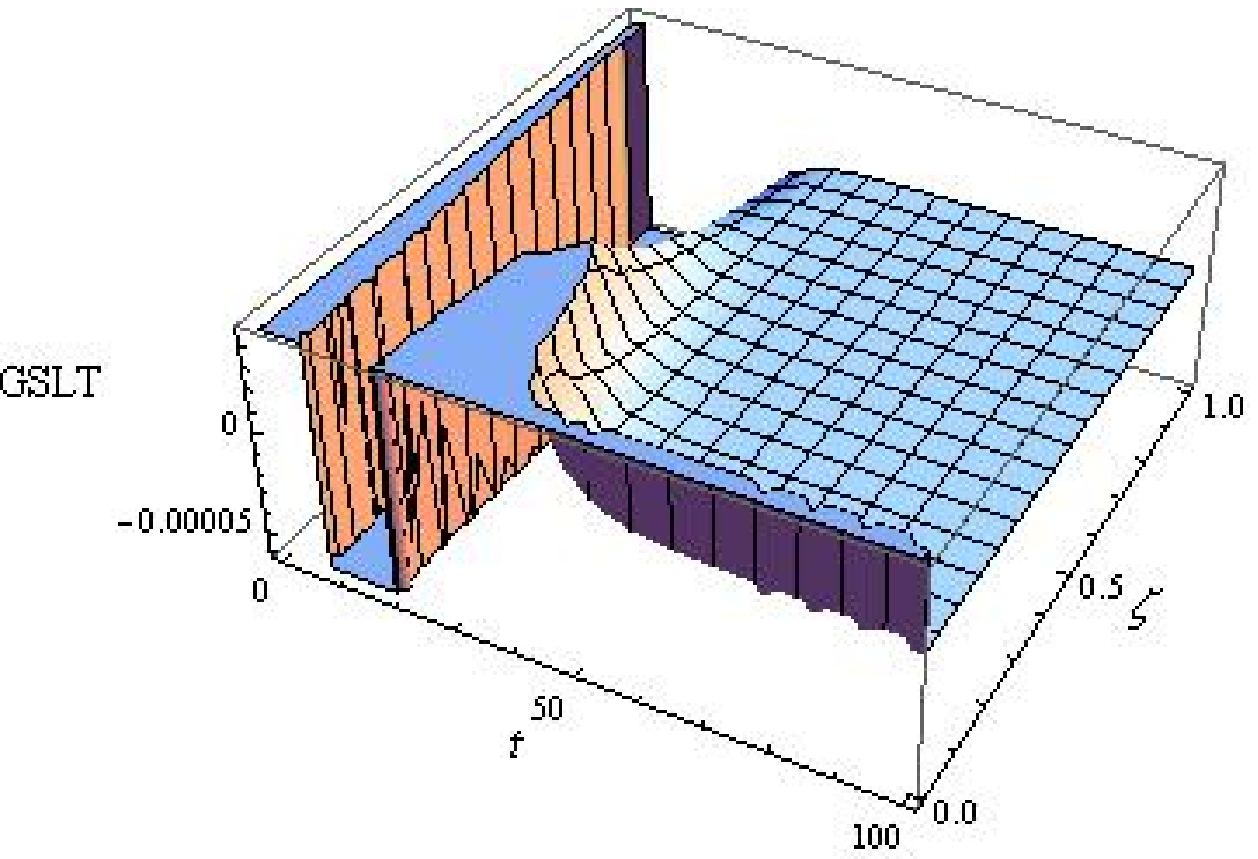,
width=0.5\linewidth}\caption{Validity of GSLT for the model
(\ref{5p}). The left plot is for $d_{1}=d_{3}=1$ and right for
$d_{1}=1$ and $d_{3}=-1$.}
\end{figure}
\begin{figure}
\epsfig{file=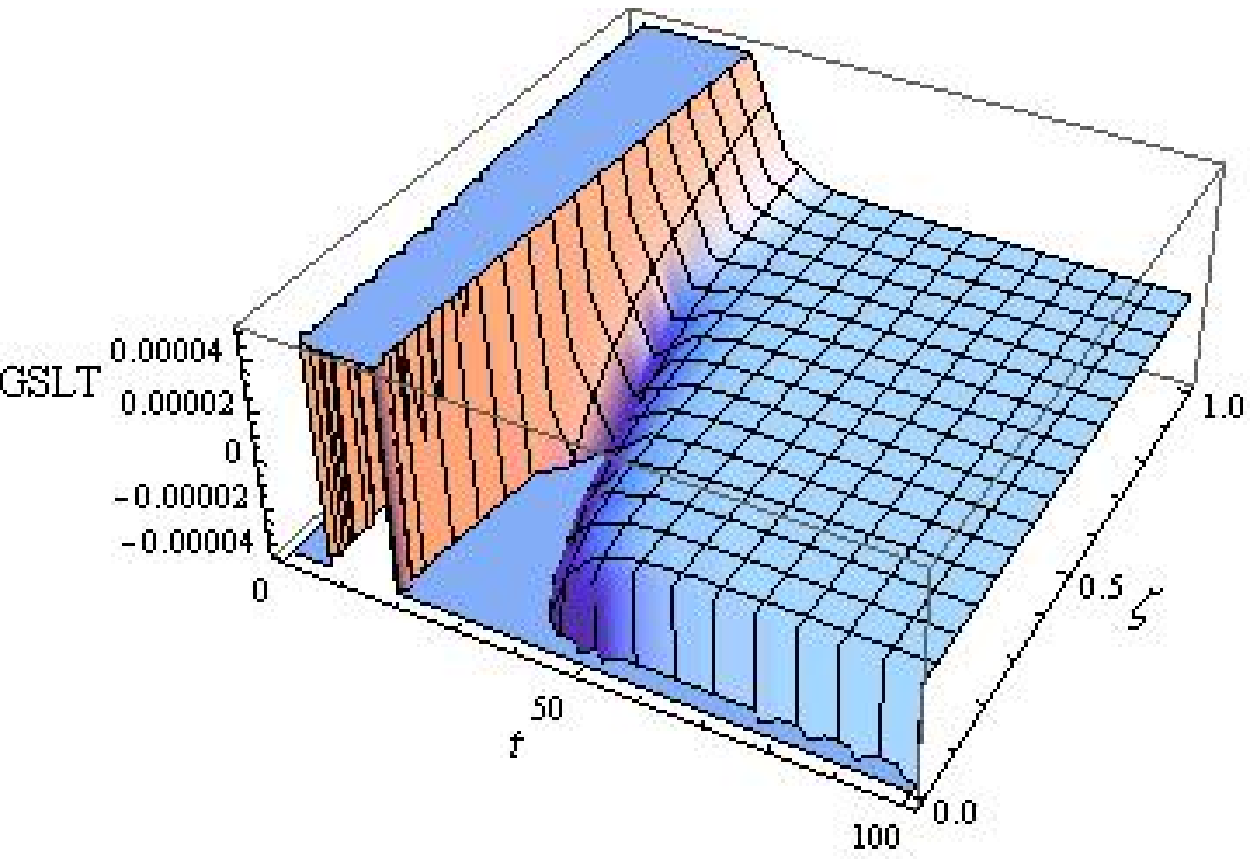, width=0.5\linewidth}\epsfig{file=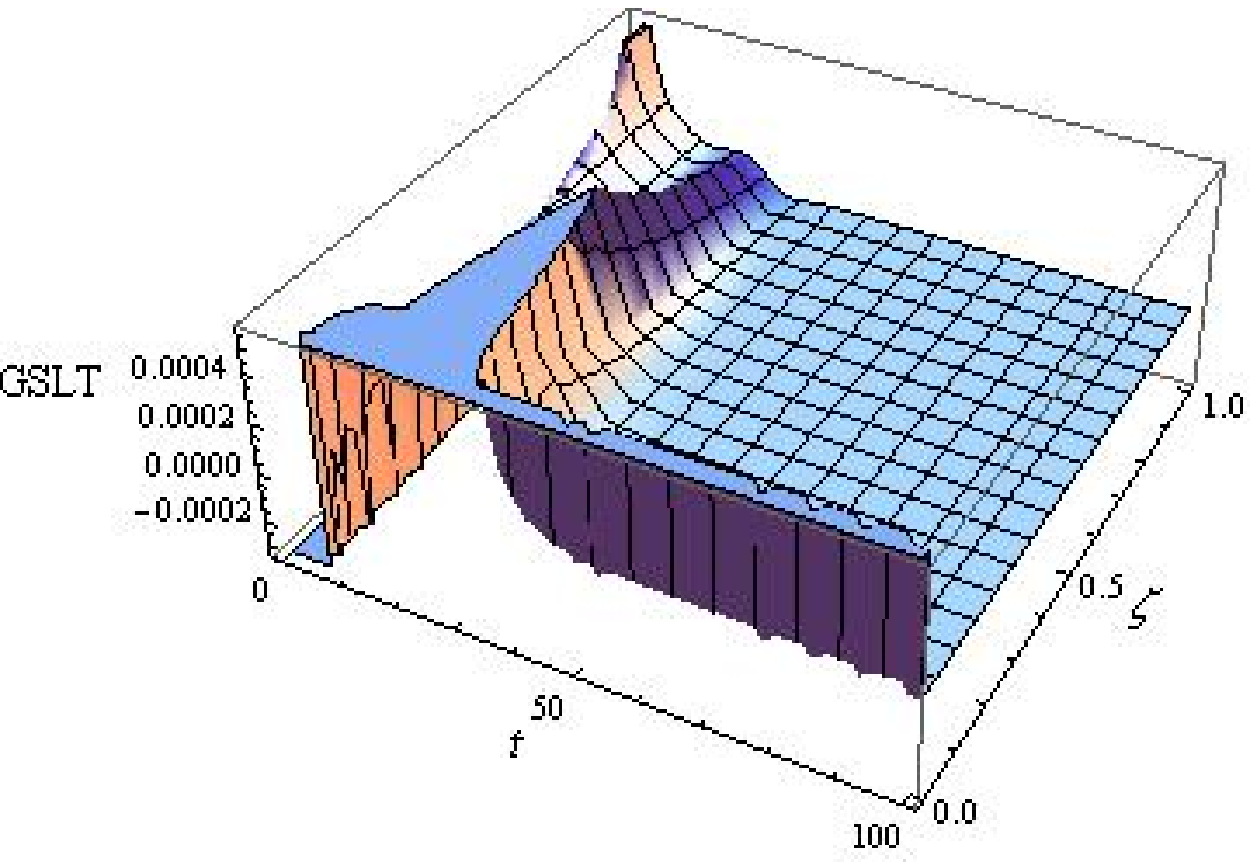,
width=0.5\linewidth}\caption{Validity of GSLT for the model
(\ref{5p}). The left plot is for $d_{1}=-1$ with $d_{3}=1$ and right
for $d_{1}=d_{3}=-1$.}
\end{figure}

The validity of GSLT for the model (\ref{5p}) is shown in Figures
\textbf{7} and \textbf{8}. For this model, the viability of law
again depends on five parameters $d_{1},~d_{2},~d_{3},~\zeta$ and
$t$ while we set $a_{0}=1,~\rho_{0}=0.3,~n=\frac{2}{3}$ and
$d_{2}=7.5$. The left panel shows that this law is satisfied for all
values of $\zeta$ at the initial times as well as when $\zeta$
approaches to $1$ with $t\geq27$ for the case $(d_{1},d_{3})>0$
while the feasible region for $d_{1}>0$ and $d_{3}<0$ is shown in
the right plot. Similarly, Figure \textbf{8} shows the regions where
GSLT holds for the remaining two signatures of $(d_{1},d_{3})$. In
this case, we observe that validity of GSLT is true for all four
possible choices of integration constants for the specific ranges of
$\zeta$ and $t$.

\section{Concluding Remarks}

In this paper, we have investigated the first and second laws in the
non-equilibrium description of thermodynamics and also checked the
validity of GSLT for reconstructed models in $f(\mathcal{G},T)$
gravity. The thermodynamical laws are studied at the apparent
horizon of FRW universe model with any spatial curvature parameter
$k$. We have found that the total entropy in the first law of
thermodynamics involves contribution from horizon entropy in terms
of area and the entropy production term. This second term appears
due to non-equilibrium behavior which implies that some energy is
exchanged between outside and inside the apparent horizon. It is
worth mentioning here that no such auxiliary entropy production term
appears in GR, GB, Lovelock and braneworld theories of gravity
\cite{12,13}.

We have found the general expression for the validity of GSLT in
terms of horizon entropy, entropy production term as well as entropy
corresponding to all matter and energy contents inside the horizon.
For non-equilibrium picture of thermodynamics, it is assumed that
temperature associated with all matter and energy contents inside
the horizon is always positive and smaller than the temperature at
apparent horizon. It is found that viability condition for this law
is consistent with the universal condition for its validity in
modified theories of gravity \cite{14}. We have also investigated
the validity condition of GSLT for the equilibrium description of
thermodynamics. The validity of this law for the reconstructed
$f(\mathcal{G},T)$ models (de Sitter universe and power-law
solution) for the dust fluid \cite{5,8} is also studied. The results
can be summarized as follows.
\begin{itemize}
\item For de Sitter reconstructed models, it is found that the
validity of GSLT is true for model (\ref{4c}) when the integration
constants $(c_{1},c_{2})$ have same signatures while for the second
model (\ref{5c}), the feasible regions are obtained for the opposite
signatures (Figures \textbf{1}-\textbf{4}).
\item For power-law reconstructed models, the valid results are
found when integration constant $d_{1}$ is positive for the model
(\ref{4p}) while for the model (\ref{5p}), this holds for all
possible choices of $d_{1}$ and $d_{3}$ (Figures
\textbf{5}-\textbf{8}).
\end{itemize}
We conclude that the validity condition of GSLT is true for both
reconstructed de Sitter and power-law $f(\mathcal{G},T)$ models with
suitable choices of free parameters.

\vspace{1.0cm}

{\bf Acknowledgment}

\vspace{0.25cm}

We would like to thank the Higher Education Commission, Islamabad,
Pakistan for its financial support through the {\it Indigenous Ph.D.
5000 Fellowship Program Phase-II, Batch-III.}\\\\
\textbf{The authors have no conflict of interest.}

\end{document}